\documentclass[11pt,english,3p,preprint,twocolumn]{elsarticle}
\usepackage[T1]{fontenc}
\usepackage[latin9]{inputenc}
\usepackage{float}
\usepackage{amsmath}
\usepackage{amssymb}
\usepackage{graphicx}

\makeatletter

\floatstyle{ruled}
\newfloat{algorithm}{tbp}{loa}
\providecommand{\algorithmname}{Algorithm}
\floatname{algorithm}{\protect\algorithmname}

\usepackage{lineno,hyperref}
\usepackage{babel}

\usepackage{amsmath,amsfonts,amsthm,amssymb}
\usepackage{graphicx}
\usepackage[percent]{overpic}

\usepackage{mathrsfs}
\usepackage{eqparbox}
\usepackage{fixltx2e}
\usepackage{stfloats}
\usepackage{pdfsync}
\usepackage{algorithmicx}
\usepackage{algorithm}
\usepackage{algpseudocode}
\usepackage{url}
\usepackage{float}
\usepackage{xcolor}
\usepackage{appendix}
\usepackage[normalem]{ulem}
\usepackage{tikz}
\usepackage{tabulary}

\journal{Ad Hoc Networks}

\newcommand\copyrighttext{Published on \emph{Elsevier Ad Hoc Networks, Volume 81, December 2018, Pages 197-210}. Available online at https://doi.org/10.1016/j.adhoc.2018.07.019 .}
\newcommand\copyrightnotice{%
\begin{tikzpicture}[remember picture,overlay]
\node[anchor=south,yshift=30pt] at (current page.south) {\fbox{\parbox{\dimexpr\textwidth-\fboxsep-\fboxrule\relax}{\copyrighttext}}};
\end{tikzpicture}%
}

\@ifundefined{showcaptionsetup}{}{%
 \PassOptionsToPackage{caption=false}{subfig}}
\usepackage{subfig}
\makeatother

\usepackage{babel}
\begin{document}

\begin{frontmatter}{}

\title{On the impact of the physical layer model on the performance of D2D-offloading
in vehicular environments}

\author{Loreto Pescosolido, Marco Conti, Andrea Passarella}

\address{Italian National Research Council,}

\address{Institute for Informatics and Telematics (CNR-IIT), Pisa, Italy\vspace*{-9mm}
}

\ead{loreto.pescosolido@iit.cnr.it, marco.conti@iit.cnr.it, andrea.passarella@iit.cnr.it}

\onecolumn
\begin{abstract}
~\vspace*{-6mm}
\\
Offloading data traffic from Infrastructure-to-Device (I2D) to Device-to-Device
(D2D) communications is a powerful tool for reducing congestion, energy
consumption, and spectrum usage of mobile cellular networks. Prior
network-level studies on D2D data offloading focus on high level performance
metrics as the \emph{offloading efficiency}, and take into account
the radio propagation aspects by using simplistic wireless channel
models. In this work, we consider a D2D data offloading protocol
tailored to highly dynamic scenarios as vehicular environments, and
evaluate its performance focusing on physical layer aspects, like
\emph{energy consumption and spectral efficiency}. In doing this,
we take into account more realistic models of the wireless channel,
with respect to the simplistic ones generally used in the previous
studies. Our objective is twofold: first, to quantify the performance
gain of the considered D2D offloading protocol with respect to a classic
cellular network, based on I2D communications, in terms of energy
consumption and spectral efficiency. Second, to show that using simplistic
channel models may prevent to accurately evaluate the performance
gain. Additionally, the use of more elaborated models allows to obtain
insightful information on relevant system-level parameters settings,
which would not be possible to obtain by using simple models. The
considered channel models range from widely used models based on deterministic
path loss formulas, to more accurate ones, which include effects like
multipath fading and the associated frequency selectivity of wideband
channels. These models have been proposed and validated, in the recent
years, through large-scale measurements campaigns. 

Our results show that the considered protocol is able to achieve a
reduction in the energy consumption of up to 35\%, and an increase
in the system spectral efficiency of 50\%, with respect to the benchmark
cellular system. The use of different channel models in evaluating
these metrics may result, in the worst case, in a sixfold underestimation
of the achieved improvement.\vspace*{-1mm}
\end{abstract}
\begin{keyword}
D2D data offloading, caching, vehicular networks\vspace*{-1mm}
\end{keyword}

\end{frontmatter}{}
\let\today\relax
~\vspace*{-10mm}

\section{Introduction\vspace*{-2mm}
}

Offloading the delivery of contents to mobile users from traditional
Infrastructure to Device (I2D) to Device-to-Device (D2D) communication,
\cite{Rebecchi2015}, ~is ~beneficial ~from ~many \hspace*{0.1mm}standpoints:\pagebreak{}~\vspace*{-4mm}

\emph{i)} Traffic congestion at the Base Station (BSs) can be reduced,
since it is no more necessary that a BS transmits the same content
to different devices that have requested it at close-by instants and
are close to each other\emph{.}

\emph{ii)} Thanks to their smaller transmission range, D2D communications
require less transmit power than I2D ones, entailing a reduction of
the overall energy consumption at the system level.
\copyrightnotice
\emph{iii) }Thanks to the increased spatial frequency reuse made possible
by the small footprint of D2D communications, with respect to I2D
ones, it is possible to increase the overall system spectral efficiency,
defined as the ratio between the amount of delivered bits and the
amount of radio resources in use, e.g., the Physical Resource Blocks
(PRBs) of a LTE-A time-frequency grid\emph{.}\vspace*{-0.5mm}

The presence of multipath fading and, in the case of wideband systems,
frequency selectivity, has a notable importance in the determination
of the transmit power used by the devices. Furthermore, by allowing
concurrent links to reuse the same radio resources, interference among
links becomes one of the major factors limiting the system performance,
and interference is highly dependent on fading and frequency selectivity.
Therefore, a careful performance evaluation of D2D data offloading
protocols, focusing on physical quantities like energy consumption
and spectrum usage, requires an accurate modeling of the radio propagation
aspects. However, most network-level studies on D2D offloading schemes,
as well as many radio resource allocation schemes for concurrent D2D
communications, assume relatively simplistic channel models. While
using these models is instrumental to simplify the analysis and devise
effective offloading algorithms, we argue that an accurate performance
evaluation should take into account more realistic channel models.\vspace*{-0.5mm}

In this work, we build upon the offloading scheme presented in \cite{Bruno2014}
and devise a Content Delivery Management System (CDMS), operated
by the network infrastructure in a distributed way. The CDMS implements
an offloading strategy suitable for dynamic scenarios like vehicular
networks. We evaluate the performance of the resulting protocol in
terms of energy consumption and spectral efficiency, by considering
realistic \emph{frequency selective} channel models. Our work extends
the analysis of the protocol presented in \cite{Bruno2014} by considering
multiple contents of interest, with different popularity. The considered
protocol falls within the class of D2D data offloading protocols (see
e.g., \cite{Whitbeck2012,Barbera2014,Rebecchi2015b,Rebecchi2016},
and \cite{Rebecchi2015} for an extensive survey). Most of the works
in this area focus on high level networking aspects and performance
metrics, while an evaluation of metrics more related to physical quantities
of interest like energy consumption and spectrum occupation, based
on accurate wireless channel models, is missing.\vspace*{-0.5mm}

The channel models considered in this work have been developed in
the last years in the context of large European projects \cite{win2chanmod2007,METIS_chanmod}
by exploiting measurement campaigns conducted by major telecom companies
like Nokia and Docomo, and have been received by standardization bodies
\cite{ITUM2135,3GPPTR36873}. To make our evaluation realistic, we
introduced in our system model an LTE-like MAC layer, with radio resources
organized in a time-frequency grid, and devised a scheduling procedure,
inspired by \cite{Yang2017}, for allocating the radio resources to
the I2D communications and D2D communications.
\vspace*{-0.5mm}

Regarding the performance gain of our offloading protocol, with respect
to a system using exclusively I2D communications, we found, for the
considered system parameters, an increase in both energy efficiency and
spectral efficiency around 50\%. Furthermore, considering the inaccuracy
in the estimation of the entity of the performance gain using simplistic
models, there can be deviations of up to one fifth (20\%), e.g.,
over an estimated value of a 32\% spectrum usage reduction (obtained using
the most accurate wireless channel model). Preliminary
results of this study have been presented in \cite{BalkanCom2017Offloading},
although under simplified assumptions for the transmission, interference,
and packet error model\footnote{In \cite{BalkanCom2017Offloading}, we assumed an ideal resource allocation,
without considering interference among concurrent D2D or I2D links,
and focused on energy consumption only.}, whereas in \cite{Pescosolido2018WoWMoM} we have devised an analytical model
to compute the offloading efficiency and energy consumption of the
CDMS considered in this work.

The paper is organized as follows: in Section~\ref{sec:Channel-modeling-approach} we provide the details
of the considered channel models. Section~\ref{sec:D2D-offloading-scheme} presents our CDMS.
Section~\ref{sec:Radio-Resource-Allocation} describes a radio resource allocation scheme on top of which the
CDMS can be implemented, and a physical layer model suitable
to capture the effect of frequency selectivity on wideband communications,
while keeping the complexity of the implementation in a system level simulator
to an acceptable level. In Section~\ref{sec:Performance-evaluation} we evaluate the performance
of the proposed offloading system, highlighting how the wireless channel
model affects the performance evaluation. Finally, in Section~\ref{sec:conclusion},
we conclude the paper, summarizing our contributions and findings.\vspace*{-4mm}

\section{Channel modeling approach\label{sec:Channel-modeling-approach}\vspace*{-2mm}
}

Many existing works on D2D-based traffic offloading involve a system-level
performance evaluation where simple channel models are used to establish
the peer-to-peer connectivity between neighboring devices, and the
interference among simultaneous transmissions on the same radio channels.
Popular models are, for instance, the ``protocol interference''
model \cite{Shi2009}, channel models with deterministic path losses,
or flat fading channels with Rayleigh fading. These models, thanks
to their simplicity, allow to evaluate high level performance metrics
in large scale networks in a relatively accurate way, but fall short
when considering metrics more directly related to physical quantities.
One high level metric, for instance, is offloading efficiency, defined
as the percentage of contents delivered through D2D communications,
\cite{Rebecchi2015}, \cite{Bruno2014}. However, while this metric
captures in a single parameter the effectiveness of the considered
offloading protocol in achieving its objective, a more in-depth performance
evaluation, which can be more easily mapped to the cost incurred by
the network operators and the users, needs to target metrics like
energy consumption and spectral efficiency. Therefore, the performance
evaluation, in this work, focuses on the latter two metrics.

When considering performance metrics such as energy consumption and
spectral efficiency, simplistic channel models are not sufficiently
accurate. In real-life deployments, shadowing, multipath small scale
fading, Line Of Sight (LOS) or Non Line Of Sight (NLOS) conditions,
induce large variations of both the useful signal and the interfering
signals strengths. For a given transmitter-receiver distance, such
variations can be in the range of tens of dBs. Furthermore, the Large
Scale Parameters (LSPs) of the random components of the channel attenuation\footnote{LSPs include, but are not limited to, first and second order moments
of LogNormal shadowing, Rician K-Factor (used to parameterize the small scale fading probability distribution),
and channel impulse response delay spread.} exhibit spatial correlation\footnote{Spatial correlation means that the LSPs of two different links with
a common transmitter (or receiver) are correlated, and the correlation
coefficient depends on the distance between the two receivers (or
transmitters).}, and correlation among different parameters for the same link. Additionally,
they are also scenario-dependent\footnote{The term ``scenario'', here, refers to the characteristics of the
real-life deployment environment, such as urban macro cell, urban
micro cell, rural, etc..} \cite{win2chanmod2007,METIS_chanmod,ITUM2135,3GPPTR36873}, and in
the same scenario they depend on the communication being either D2D
or I2D. Finally, in existing wideband systems, frequency selectivity
due to multipath fading is also an important physical layer aspect that
needs to be taken into account for an accurate performance evaluation.

In this work, motivated by the multitude of physical effects overlooked
by widely used simplistic modes, we adopt the geometry-based stochastic
channel model (GSCM) put forward by the WINNER II European project
in \cite{win2chanmod2007} and subsequently refined/extended by the
ITU \cite{ITUM2135}, the 3GPP \cite{3GPPTR36873}, and by the METIS
Project \cite{METIS_chanmod}. Specifically, in \cite{ITUM2135},
a detailed procedure is described for generating a set of frequency
selective channels affected by shadowing and small scale fading. The
procedure uses specific, scenario-dependent, formulas (that are provided
by the above referenced reports), for computing the path loss and
generating the set of LSPs.

In order to develop a large scale network level simulator, able to
cope with tens or hundreds of nodes, we implemented the detailed channel
models developed by the WINNER II and METIS Projects \cite{win2chanmod2007,METIS_chanmod,ITUM2135,3GPPTR36873},
although skipping some details, like the intra-path superimposition
of micro-paths, which would increase the complexity and memory usage
of the simulator to an unnecessary (for our purposes) level. Our
implementation is based on a discretized 2D representation of the
Region Of Interest (ROI). It first involves the computation of the
path loss between any two points in a 2D rectangular grid with a spatial
step of $5\,\text{m}$ in both dimensions, and between each Base Station
(BS) and each point in the grid. To do this, we used the scenario-dependent
formulas recommended in \cite{METIS_chanmod} for the Urban micro-cell
(UMi) and Urban macro-cell (UMa) scenarios. Then, random correlated
LSPs are generated for each link. Finally, the set of multipath channels,
including random amplitude and delay for each signal replica that
arrives at the receiver (through the multiple paths), is generated.
This is done according to the procedure described in \cite{win2chanmod2007}
for the GSCM model. While the procedure we used is the one presented
in \cite{win2chanmod2007}, we parameterized all the formulas to generate
the LSPs and the impulse responses with the updated parameter values
proposed in \cite{METIS_chanmod}, for the UMi and UMa scenarios,
and for both I2D and D2D communications \footnote{ Notably, \cite{METIS_chanmod} also includes parameter sets specific
of D2D communications (which differ from I2D ones in that transmitter
and receiver, in the D2D case, are typically at the same height, which
is not the case for I2D communications).}. In this work, we assume omnidirectional antennas and account for
the different heights of the transmitting antenna for I2D and D2D
communications (which do have an effect on the path loss values).
Specifically, we consider an antenna height of 10~m and 25~m for
BSs (for the UMi and UMa case, respectively) and $1.5\,\text{m}$
for mobile devices, as recommended in \cite{METIS_chanmod,ITUM2135}.
The models in \cite{win2chanmod2007,ITUM2135,3GPPTR36873,METIS_chanmod}
provide a further level of detail, which we decided to skip as they
are not relevant to our scenario and would unnecessarily increase
the complexity of the simulator\footnote{Most of the details we skipped become relevant when dealing with directional
antennas, multi-element antennas, and MIMO communications, which we
do not consider in this work.}. \vspace*{-4mm}

\section{D2D data offloading scheme with delayed content delivery\label{sec:D2D-offloading-scheme}\vspace*{-2mm}
}

In D2D Data offloading schemes, D2D communications occur as mobile devices, although they can obtain
a desired content by the infrastructure, i.e., the BSs, would preferably
receive it from neighboring devices, that have cached the content
previously. Offloading protocols require that a mobile device, upon
receiving a content, keeps it cached, for some time, in order to share
it with neighbors which will request it within the amount of time
it is kept in the cache.

In static or quasi-static scenarios, a content request from a device
is typically fulfilled immediately: either at the time of request
there is a device (close-by the requesting device) with the content
in its cache, which shares it with the requesting device, or, if there
is no such neighboring device, the requesting device receives the
content from a BS. However, in highly dynamic scenarios, such as in vehicular
environments, thanks to the nodes' mobility, even though a certain content
may be unavailable from the neighbors of the requesting device at
the request time, it could still be obtained through a D2D link, later
on, from a \emph{new} neighboring device, which has previously cached
the content. The new neighboring device is just a device that has
come close to the requesting device \emph{after} the content request
time. To exploit this possibility, the offloading strategy presented
in \cite{Bruno2014}, uses a \emph{content timeout}, i.e., an interval
starting at the content request instant, during which the requesting
node tries to obtain the content from the devices it encounters. If
this is not the case, the content is delivered to the node by a BS,
at the expiration of the content timeout. In this way, the system
offloading efficiency can be increased, with respect to the choice
of transmitting the content immediately, either by a neighbor (if
available) or by a BS. Typically, this approach can be taken for contents
related to non delay-critical applications for which it is reasonable
to accept that the content, instead of being retrieved immediately,
can be obtained in a matter of seconds, or tens of seconds, depending
on the application. Moreover, this approach works also when content
requests are not synchronised, and therefore the BS cannot serve multiple
mobile nodes through multicast transmissions.\vspace*{-1mm}

In this work, we propose a CDMS that controls this type of offloading system. The CDMS is implemented
as a distributed software agent that can communicate and track the
users along their paths through the ROI. The proposed CDMS executes
a protocol that implements the same offloading strategy of \cite{Bruno2014}.
However, whereas in \cite{Bruno2014} the implementation of the offloading
strategy is distributed at the device level, our CDMS mostly relies
on control operations executed, still in a distributed way, at the
network infrastructure nodes. Hence, whereas \cite{Bruno2014} largely
relies on the self-organization capabilities of the devices, our protocol
implements an infrastructure-assisted D2D communication approach.
The motivation is that, in highly dynamic scenarios, the signaling
required by the neighbor discovery protocol and content requests notifications
of completely decentralized solutions may be excessive, if performed
exclusively by the nodes, given that the network topology is continuously
changing and its representation needs to be kept up to date.\vspace*{-3mm}

\subsection{Content delivery management\label{sec:Content-Delivery-Management}}

We consider a ROI populated with mobile users (humans, vehicles) each
carrying a wireless device. Specifically, we assume that mobile devices
are on board of vehicles. The devices may be either human hand-held
devices whose owners are in the vehicles, or wireless devices installed
in the vehicle equipment. Users enter, roam into, and exit the ROI.
As a vehicle enters the ROI, its on board device starts requesting
contents according to a given content request process\footnote{The content request process is originated at the application layer.
Here, it is of no importance whether the interest is generated by
a human or by, for instance, an IoT application executed by the software
on a vehicle.}. The content request process is characterized by a content-request
\emph{arrival} process, which defines the time instants at which the
requests are generated, and a content-interest probability distribution,
which defines which content is requested. We assume that the requests
instants of different devices are statistically independent, and that
the specific contents requested by different devices, or by the same
device in different requests, are also independent. In Section~\ref{sec:Performance-evaluation},
we provide the specific models for the nodes' mobility and the content
request process used in our simulations. However, our protocol does
not require the knowledge of these models. 

During its path within the ROI, at each instant, each device is associated
to a BS, which is responsible of handling the delivery of the contents
requested by that device. For each device $k$ associated to a BS,
the BS holds a list of neighbors $\mathcal{N}_{k}$ composed of pairs
of the form $(j,r_{j}^{k})$ , where $j$ is the id of any node which
is a neighbor of node $k$, and $r_{j}^{k}$ is a ranking index of
node $j$ as ``seen'' by node $k$ on the basis of a given criterion.
In this work, the criterion to establish if two nodes are neighbors,
and the ranking of each node's neighbors, is based on a nominal indicator
of the channel quality between the nodes. In general, the nominal
channel quality may be computed, by the CDMS, on the basis of the
positions of the nodes, which the BSs are assumed to know, using a
given channel model. In the following. we indicate this measure as
\emph{nominal channel gain.} Assuming a deterministic model for the
computation of the \emph{nominal} channel gain (i.e., that the nominal
channel gain $g$ is a monotonically decreasing function of the distance
$d$), two nodes are considered as neighbors, by the CDMS, in the
time instants their distance is less then or equal to a maximum value
$d_{\max}$. In this work, we assume that a deterministic model is
used to determine if two nodes are neighbors\footnote{More sophisticated techniques may be considered to determine the nominal
channel gain. For instance, it may be computed taking into account
slowly varying shadowing statistics between any to points in the (discretized)
space. This would require means to create such a shadowing map. This
is an interesting research issue which we will will consider in future
works.}.

The CDMS, essentially, acts on a distributed database containing the
up to date list of each node's position, the list of its neighbors,
and the nominal channel gain between any two neighbors and between
each node and the surrounding BSs. The CDMS can be implemented as
a distributed software agent, for instance, through a Network Function
Virtualization (NFV) component allocated appropriately in the infrastructure
topology. The BSs are used by the CDMS to operate the offloading protocol.
 To handle the handover of ongoing requests originated from a node
that crosses a cell border while waiting for a content, adjacent BSs
periodically exchange the up to date status of the ongoing request
procedures (see below) of the nodes moving across cells\footnote{This information exchange can be performed using high speed fiber
connections, or dedicated radio channels forming a wireless mesh-type
backhaul component of the Radio Access Network (RAN).}. At each BS, the lists $\mathcal{N}_{k}$ are updated on the basis
of Hello messages, sent periodically by the nodes, containing their
id. Each node $k$ has an internal content cache $\mathcal{C}_{k}$
populated with previously downloaded contents. At any time, the CDMS
also has an index of the contents in each node's cache, although the
CDMS does not necessarily hold a copy of the contents.\vspace*{-0.5mm}

Based on this available information, following a content request by
device $k$ for content $z$, the CDMS determines the content delivery time and mode.
Three cases are possible:\vspace*{-0.5mm}

1) At the request time, node $k$ has at least one neighbor with the
content $z$ available. In this case, the CDMS prompts the closest
neighbor with content $z$ available to transmit it to node $k$.\vspace*{-0.5mm}

2) At the request time, node $k$ has no neighbors with the content
$z$ available, but during the content timeout, as a consequence of
the mobility of the devices, a device with the content available comes
within the range $d_{\max}$ off the node $k$. In this case, as soon
the CDMS detects that such condition is verified, it prompts the node
that has come within the range $d_{\max}$ to immediately transmit
the content to node $k$.\vspace*{-0.5mm}

3) No neighbor of node $k$, either at the request time, or during
the content timeout, has content $z$ in cache. In this case, at the
end of the content timeout, the CDMS prompts the transmission of content
$z$ by the infrastructure, i.e., by the closest BS.

The pseudocode in Algorithms 1 and 2 describes in more detail the
actions taken on demand, i.e., as a consequence of content requests,
by the nodes and the CDMS. We briefly introduce the notation required
for a correct interpretation of the algorithms: the notation $\biguplus\left\{ \mathcal{C}_{j}|\text{condition on }j\right\} $
is used to indicate the union of the caches of nodes satisfying a
given condition; the notation $\hat{j}(k,z)$ is used to indicate
the node $j$ that has the best ranking $r_{j}^{k}$ among the neighbors
of node $k$ which have content $z$ in their caches; the notation
$j\stackrel{z}{\rightarrow}k$ indicates the transmission of content
$z$ from node $j$ to node $k$. These transmissions are triggered
by the CDMS. The remaining notation used in Algorithms 1-2 is self-explaining. 

Algorithm 1 describes the actions of a node as it becomes interested
in a content. Essentially, it notifies the CDMS that it is interested
in that content, and then waits for receiving it either from a BS
or from a neighbor. The system guarantees that the content will be
delivered within a given \emph{content timeout}. After the reception
of the content, the node makes it available for other nodes that may
request it, for a limited amount of time determined by a \emph{sharing
timeout}. The sharing timeout is required to avoid cache overflow.
\begin{algorithm}[t]
\caption{{\small{}Actions taken by node $k$ to request content $z$$\protect\phantom{()}$}}
\label{algo:node_req_handle}

\footnotesize
\begin{enumerate}
\item $\mathbf{Upon}$ request for content $z$ from the application layer
\item $\mathbf{Set}$ \texttt{$k$\_content\_received} = $\mathbf{false}$
\item $\mathbf{Send}$ $(k,z)$\texttt{\_cont\_req} to CDMS
\item $\mathbf{while}$ \texttt{$k$\_content\_received ==} $\mathbf{false}$
$\mathbf{do}$\\
~\hspace*{\fill}$\triangleright$ Wait for receiving content $z$,
from a BS or from a neighbor
\item \quad{}$\mathbf{if}$ content $z$ is received $\mathbf{then}$
\item \quad{}\quad{}$\mathbf{Set}$\texttt{ $k$\_content\_received} =
$\mathbf{true}$
\item \quad{}\quad{}$\mathbf{Send}$ $(k,z)\_$ACK to CDMS and/or the
sending node
\item \quad{}\quad{}$\mathbf{Add}$ $z$ to $\mathcal{C}_{k}$
\item \quad{}\quad{}$\mathbf{Set}$ \texttt{$(k,z)$\_sharing\_timeout}
\item \quad{}\quad{}$\mathbf{break}$
\item \quad{}$\mathbf{end\,if}$
\item $\mathbf{end\,while}$
\item $\mathbf{while}$ \texttt{$(k,z)$\_sharing\_timeout} is not expired
$\mathbf{do}$\\
~\hspace*{\fill}$\triangleright$ Available for opportunistic sharing
of content $z$
\item \quad{}$\mathbf{Upon}$ request from CDMS (step 7 of Algorithm \ref{algo_CDMS_req_handle})
\item \quad{}$\mathbf{Send}$ $z$ to node requesting it
\item $\mathbf{end\,while}$
\item $\mathbf{Remove}$ content $z$ from $\mathcal{C}_{k}$
\item $\mathbf{Cancel}$ \texttt{$(k,z)$\_sharing\_timeout}
\end{enumerate}
\end{algorithm}

Algorithm 2 describes the actions taken by the CDMS to handle a content
request. Here, a key point, which allows to further reduce the system
energy consumption, is that the CDMS selects the best node for delivering
the content, on the basis of channel quality considerations, represented
by the ranking of each node's neighbors (steps 5-13). If, however
the content cannot be delivered through a D2D communication within
the content timeout, the CDMS uses the BSs to deliver it (steps 15-20).

\section{Radio resource allocation, transmission, and error model\label{sec:Radio-Resource-Allocation}}

\subsection{In-band radio resource reuse}

To evaluate the energy consumption and the spectral efficiency of
our protocol, we implemented a MAC layer structure exploiting the
time-frequency domain, and designed a scheduling and resource allocation
protocol that handles the transmission of contents through either
D2D or I2D communications. The BSs schedule both the transmission
of I2D and D2D packets with the periodicity of a control interval
(CI) of duration $T_{\text{CI}}$, using an overall system bandwidth
$W$.
\begin{algorithm}[t]
\caption{{\small{}Actions taken by CDMS for handling content request $(k,z)$}}
\label{algo_CDMS_req_handle}

\footnotesize
\begin{enumerate}
\item $\mathbf{Upon}$ receiving \texttt{$(k,z)$\_cont\_req}
\item $\mathbf{Set}$ \texttt{$(k,z)$\_served} = $\mathbf{false}$
\item $\mathbf{Set}$ \texttt{$(k,z)$\_content\_timeout} 
\item $\mathbf{while}$ \texttt{$(k,z)$\_content\_timeout} is not expired
$\mathbf{do}$
\item \quad{}$\mathbf{if}$ \texttt{$z\in\biguplus\left\{ \mathcal{C}_{j}|j\in\mathcal{N}_{k}\right\} $}
$\mathbf{then}$
\item \quad{}\quad{}$\mathbf{Identify}$ $\hat{j}(k,z)$
\item \quad{}\quad{}$\mathbf{Trigger}$ transmission $\hat{j}(k,z)\stackrel{z}{\rightarrow}k$
\item \quad{}\quad{}$\mathbf{Wait}$ for \texttt{$(k,z)$\_}ACK
\item \quad{}\quad{}$\mathbf{Upon}$ \texttt{$(k,z)$\_}ACK reception
\item \quad{}\quad{}$\mathbf{Set}$ \texttt{$(k,z)$\_served} = $\mathbf{true}$
\item \quad{}\quad{}$\mathbf{Remove}$ $(k,z)$ from $\mathcal{L}_{\text{req}}$
\item \quad{}\quad{}$\mathbf{break}$
\item \quad{}$\mathbf{end\,if}$
\item $\mathbf{end\,while}$
\item $\mathbf{if}$ \texttt{$(k,z)$\_served} == $\mathbf{false}$
\item \quad{}$\mathbf{Send}$ \texttt{$z$} to \texttt{$k$}
\item \quad{}$\mathbf{Wait}$ for ACK\texttt{\_}$(k,z)$
\item \quad{}$\mathbf{Upon}$ reception of ACK\texttt{\_}$(k,z)$
\item \quad{}$\mathbf{Set}$ \texttt{$(k,z)$\_served} = $\mathbf{true}$
\item $\mathbf{end\,if}$
\item $\mathbf{Cancel}$ \texttt{$(k,z)$\_content\_timeout}
\end{enumerate}
\end{algorithm}
In each CI, the radio resources are organized in a grid over a time-frequency
frame of duration $T_{\text{CI}}$ and bandwidth $W$. Each grid element
is a Physical Resource Block (PRB) of duration $\tau$ and bandwidth
$w$. We indicate the number of PRBs in a CI as $N_{\text{PRB}}$.
The bandwidth of a PRB contains $n_{c}$ subcarriers, each of bandwidth
(or subcarrier spacing) $w_{c}=w/n_{c}$. For each CI, a fine grain
scheduling procedure has to be executed, which determines which PRBs
will be used by all the involved nodes to transmit the packets that the
CDMS has selected for transmission in that CI. To distinguish it from
the scheduling performed by the CDMS, which acts on a coarser time
scale, in the order of the content timeout duration, we call Radio
Resource (RR) scheduler the scheduler that, in each CI, effectively
allocates PRBs to each link (and for each link, to the packets that
have to be transmitted over it). The RR scheduler is not the focus
of this work, but serves as a baseline implementation to analyze the
performance of our CDMS in a realistic setting. Therefore, for the
sake of readability, we provide here a high level description, while
a more detailed description is presented in the Appendix.

To leverage spatial frequency reuse, we have devised a RR scheduler
inspired by the ``full resource sharing'' approach of \cite{Yang2017}.
With this approach, multiple concurrent D2D communications can share
the same PRBs, and each PRB can be assigned also to, at most, a single
I2D link within a cell and across adjacent cells. The considered approach
is particularly attractive since it allows to exploit spatial frequency
reuse to its extreme, outperforming other conventional approaches
like user-oriented resource sharing or resource-oriented resource
sharing \cite{Yang2017}. For the purpose of this work, however, the
RR scheduler proposed in \cite{Yang2017} cannot be directly applied,
since it is devised under the assumptions of a single cell, uniform
transmit power for all D2D communications and for all I2D communications,
and flat fading channels. Additionally, in \cite{Yang2017}, resources
are allocated to links with the granularity of a single radio resource
(for us, a PRB), with the objective to maximize throughput. With
our setup, in which there is a predefined content to be transmitted
by each scheduled link in each CI, this could result in a link receiving
an amount of radio resources which is not sufficient to transmit the
desired content within a CI\footnote{For simplicity, our MAC does not consider packet fragmentation over
multiple CIs.}. Therefore, our radio resource allocation, although inspired by the
one presented in \cite{Yang2017}, differs from it in several aspects.
Our implementation of the RR scheduler includes an admission
control which limits the number of both I2D and D2D packets that can
be transmitted using the resources in a CI, based on an iterative
validation of cross-link interference constraints (see the Appendix).
If a packet transmission cannot fit, it is pruned, and postponed to
the next CI. These episodes may occur, for instance, during localized
traffic load peaks. Note that, in general, the latencies introduced
due to pruning, and to transmission errors (with the associated retransmissions)
are negligible with respect to the order of magnitude of the delay-tolerance
(tens of seconds), i.e., the content-timeout\footnote{Note that the content timeout parameter may also be set in a conservative
way, e.g., 1 or 2 seconds less than the actual delay tolerance.}. Delays due to transmissions errors and pruning, however, are actually reproduced in our simulations.

In the following, we indicate with $S$ the set of radio links that are scheduled for
transmission in a given CI. A link $i$ represents a transmitter-receiver
pair $(x_{i},y_{i})$. In our setup, $x_{i}$ can be either a BS or
a mobile device, whereas $y_{i}$ is always a mobile device. Let $k$
be the generic PRB in a CI, $k\in\{1,...,N_{\text{PRB}}\}$, and $p$
an index over the packets that are scheduled to be transmitted during
the CI. The output of the RR scheduler can be represented by the
set of indicator functions $\phi_{i,k}(p),\,\forall i\in S,\,\forall k\in\{1,...,N_{\text{PRB}}\}$,
which tells that link $i$ uses PRB $k$ to transmit packet $p$,
i.e.,\vspace*{-2mm}
\begin{equation}
\phi_{i,k}(p)=\begin{cases}
1 & \text{if PRB}\,k\,\text{is used by link}\,i\\
 & \text{to transmit content }p\\
0 & \text{otherwise}
\end{cases}.\label{eq:PRBs_allocation}
\end{equation}

\subsection{Physical layer and transmission error model\label{subsec:Transmission-and-error}\vspace*{1mm}
}

In the following, we describe a physical layer model which enables
to perform large scale network-level simulations while retaining (to
an acceptable extent) the representation of the effect of important
physical layer parameters that are tightly related to the wireless
channel model: shadowing, fading, and frequency selectivity.

We assume a multicarrier modulation with a uniform power allocation
over all the subcarriers in use. We denote with $P_{i}^{(c)}$ the
transmit power used by the transmitter of link $i$ on each subcarrier
it uses, and with $H_{\left\langle j,i\right\rangle }\left(f\right)$
the channel transfer function of the radio channel between the transmitter
of link $j$ and the receiver of link~$i$. Considering the PRB $k$,
we enumerate its $n_{c}$ subcarriers as $f_{1}^{(k)},\ldots,f_{n_{c}}^{(k)}$.
The Signal to Noise Plus Interference Ratio (SNIR) for link $i$ in
the $q$-th subcarrier of PRB $k$ is
\begin{equation}
\xi_{i,q}^{(k)}=\frac{P_{i}^{(c)}\left|H_{\left\langle i,i\right\rangle }\left(f_{q}^{(k)}\right)\right|^{2}}{\sigma_{c}^{2}+\sum_{j\in S\backslash\{i\}}\psi_{j,k}P_{j}^{(c)}\left|H_{\left\langle j,i\right\rangle }\left(f_{q}^{(k)}\right)\right|^{2}},\label{eq:SNIR_f}
\end{equation}
where $\sigma_{c}^{2}$ is the noise power on each subcarrier. Specifically,
$\sigma_{c}^{2}=w_{c}FN_{0}$, where $N_{0}$ is the thermal noise
power spectral density, and $F$ is the noise figure of the receiver.

Our error model for the transmission, on link $i$, of a content $p$
of size $L(p)$ payload bits, using the PRBs allocated to its transmission
by the RR scheduler, consists in verifying that the corresponding
\emph{achievable amount of information}\footnote{We use the term ``achievable amount of information''
of a transmission over a subcarrier, in an information theoretic sense.
Specifically, we define it as the product of a Shannon capacity, haircut
by a maximum transmit spectral efficiency (i.e. a given maximum number
of bits that can be loaded on each symbol of the digital modulation
scheme in use), times the duration of the transmission.} transferrable by link $i$ in those PRBs is larger than or equal
to $L(p)$. 

The Shannon capacity of a single subcarrier, normalized to its bandwidth,
is given by $\log_{2}\hspace{-1mm}\left(1+\xi_{i,q}^{(k)}\right)$,
where $\xi_{i,q}^{(k)}$ is given by \eqref{eq:SNIR_f}. Let $e_{i}$
be the \emph{transmit spectral efficiency} (i.e., the maximum
number of bits that can be carried by a single symbol of the digital
modulation scheme in use, over a subcarrier) of link $i$, measured
in bits per second per Hertz (bps/Hz). In real systems, a subcarrier
use cannot carry an amount of information larger than $e_{i}$, even
when its Shannon capacity is larger than it. Therefore, we compute the contribution
of, say, the $q$-th subcarrier used in a given PRB $k$ (i.e., for the duration of a time slot), to the achievable amount of information, as
$\tau w_{c}\min\left(e_{i},\log_{2}\hspace{-1mm}\left(1+\xi_{i,q}^{(k)}\right)\right)$.
Summing up all the contributions of this kind, on the basis of the
PRBs allocated to the transmission of a packet $p$ (indicated by
the function $\phi_{i,k}^{(p)}$) by the RR scheduler, and imposing
that the sum is larger than the amount of \emph{payload} bits, we
obtain the following inequality, with which we model a packet transmission
success:
\begin{equation}
\small{\tau w_{c}\sum_{k=1}^{N_{\text{PRB}}}\phi_{i,k}^{(p)}\sum_{q=1}^{n_{c}}\min\left(e_{i},\log_{2}\hspace{-1mm}\left(1+\xi_{i,q}^{(k)}\right)\right)\hspace{-1mm}\geq\hspace{-1mm}L(p).}\label{eq:successful-tx}
\end{equation}

If \eqref{eq:successful-tx} is satisfied, the packet is successfully
received, otherwise, we count a failed transmission.

Up to now, we have left unspecified the settings of two important
parameters: the transmit power $P_{i}^{(c)}$ appearing in Eq.~\eqref{eq:SNIR_f},
and the number of PRBs to be allocated to a packet transmission, i.e,
the number of nonzero terms in the double sum of the left hand side
of \eqref{eq:successful-tx}. The latter parameter is determined by
the need to encode each content in a packet that has a size, $D(p)$,
which is, in general, larger than $L(p)$. We discuss these two parameters
in the following subsection.

\subsection{Physical layer parameters: link margin and forward error correction\label{subsec:Forward-error-correction}}

In the practical deployment of any wireless network which uses power
control based on nominal channel gains, the transmitter needs to
compensate for the presence of random shadowing and multipath fading. In our case, 
these random effects are not factored in by the nominal channel gain $g_{\left\langle i,i\right\rangle }$
(on which $P_{i}^{(c)}$ depends). Moreover, in the presence of concurrent
links using the same radio resources, the effect of the interference
coming from the links using the same PRBs must also be taken into account.
The typical tools to deal with these problems, are: i) a transmit
power link margin, added on top of the power that would be required
for transmitting over an ideal deterministic channel with flat frequency
response, ii) Forward Error Correction (FEC), or, more in general,
Adaptive Modulation and Coding (AMC) (see below). 

We model the use of FEC as follows: we consider a coding rate parameter,
which we indicate as $K_{\text{ec}}$, defined as the ratio between
the payload size $L$ and the coded packet size $D$, or $K_{\text{ec}}=L/D$. $K_{\text{ec}}$ is a characteristic
figure of the FEC code in use, which tells how
much redundancy is introduced in the coded packets to increase the
resiliency to noise and interference. The lower $K_{c}$, the higher
the redundancy, at the price of a larger number of bits to be transmitted,
for a given payload. The selected coding rate (and hence number of
\emph{coded} bits $D$), in our model, determines the number of nonzero
elements in the right hand side of \eqref{eq:successful-tx}, i.e.,
those for which $\phi_{i,k}^{(p)}=1$ ). The \emph{specific} elements
for which $\phi_{i,k}^{(p)}=1$ are the output of the RR scheduler.

Considering an AMC system, the power allocated over a subcarrier
(excluding the power margin) would be set to the power required to
support a target normalized information rate equal to a transmit spectral
efficiency selected within a discrete set $\{e_{\min},\ldots,e_{\max}\}$,
representing, e.g., the use of different digital symbols constellations.

We set the power $P_{i}^{(c)}$ appearing in Eq.~\eqref{eq:SNIR_f}
as\vspace*{-2mm}
\begin{equation}
P_{i}^{(c)}=M\frac{\sigma_{c}^{2}}{g_{\left\langle i,i\right\rangle }}\left(2^{\overline{e}_{i}}-1\right),\label{eq:power_control}\vspace*{-1mm}
\end{equation}
where $g_{\left\langle i,i\right\rangle }$ is the nominal channel
gain of link $i$, $\overline{e}_{i}$ is a target normalized information
rate (measured in bps/Hz) that the communication aims to achieve,
and $M$ is a suitable link margin. The term $\frac{\sigma_{c}^{2}}{g_{\left\langle i,i\right\rangle }}\left(2^{\overline{e}_{i}}-1\right)$
is the transmit power per subcarrier that would be required to support
the desired normalized information rate, $\overline{e}_{i}$, over
an ideal link with a flat fading channel whose gain is $g_{\left\langle i,i\right\rangle }$.
In dBm, the per-subcarrier transmit power is given by\vspace*{-1mm}
\begin{align}
P_{i,\text{[dBm]}}^{(c)}= & \sigma_{c,\text{[dBm]}}^{2}-g_{\left\langle i,i\right\rangle \text{[dB]}}+10\log_{10}\left(2^{\overline{e}_{i}}-1\right)\nonumber \\
 & +M_{\text{[dB]}.}\label{eq:Power_per_subcarrier_dBm}\vspace*{-1mm}
\end{align}

The above described physical layer model is able to reproduce the
behavior of (coexisting) communications occurring on wideband frequency
selective channels, taking into account the power link margin and
AMC parameters like FEC coding rate and constellation size, without
the need to simulate the transmission at the symbol level (which would
make it quite hard to perform system level simulations involving hundreds
of nodes). Furthermore, the RR scheduler we devised based on \cite{Yang2017},
can exploit different settings of the said parameters (link margin,
FEC coding rate, and digital symbol constellations).

In the rest of this work, since the focus is on the effect of channel
modeling on the performance evaluation, rather then on finding the
optimal AMC parameters and power margin balance, we consider the 
FEC coding rate and transmit spectral efficiency as given parameters,
i.e., we consider a single $e_{i}$ and set $\overline{e}_{i}=e_{i}$
in the transmit power expression \eqref{eq:power_control}. We focus
then on the power margin setting, which, as described below, is a
function of the wireless channel model.

Consider an interference-free link: in the presence of shadowing,
flat or frequency selective fading, without adding a link margin (i.e,
setting $M_{\text{[dB]}}=0$ in \ref{eq:Power_per_subcarrier_dBm}),
the achievable normalized information rate on a given subcarrier,
given by $\log_{2}\hspace{-1mm}\left(1+\xi_{i,q}^{(k)}\right)$, may
drop below $\overline{e}_{i}$ for a considerable percentage of the
subcarriers (even with a null interference term in the denominator
of \eqref{eq:SNIR_f}). The achievable amount of  information
in Eq.~\eqref{eq:successful-tx} is a random variable with a specific
probability density function, which depends on the channel model in
use. The link margin $M$ is therefore set to guarantee a prescribed
outage probability, i.e., that Eq.~\eqref{eq:successful-tx} is \emph{not}
satisfied with a probability lower than a given threshold, $p_{\text{tx,out}}\ll1$.
It is clear that the suitable $M$ critically depends on the radio
propagation aspects and frequency selectivity of the channels in a
given deployment scenario and/or channel model. To take into account
the link margin in our performance evaluation, we computed, through
montecarlo simulations, the link margins required to guarantee an
outage probability below 0,5\%, assuming a FEC coding rate $K_{\text{ec}}=0,8$,
for the three channel models (out of the six channel models considered
in this work, see Section~\ref{subsec:Channel-models-discussion}) that present a stochastic component. As expected, the
required link margins are different for the three stochastic channel
models. The specific values are reported in the next section.\vspace*{-4mm}

\section{Performance evaluation\label{sec:Performance-evaluation}}\vspace*{-2mm}

Before presenting our performance evaluation results, we preliminarily
discuss, in the following subsection, the most relevant channel models
features that may affect the performance evaluation accurateness.
In Subsection~\ref{subsec:Considered-scenarios} we describe the
considered deployment scenarios, and in Subsection~\ref{subsec:Simulation-results}
we present our simulation results.\vspace*{-3mm}

\subsection{Channel models and implications\label{subsec:Channel-models-discussion}}

To describe the different models, let $P_{tx}$ be the transmit power
per subcarrier used by a transmitter, and $P_{rx}(d)$ the power received
at a distance of $d$ meters. In the following, all power values are
expressed in mW or in dBm\footnote{In the latter case ``dBm'' appears as
a subscript.}. $PL_{\text{dB}}$ indicates the path loss in decibels
($PL_{\text{dB}}=P_{tx,\text{dBm}}-P_{rx,\text{dBm}}$). We consider
the following six different channel models, enumerated as M1-M6. The
considered models are also summarized in Table~\ref{tab:chanmod}.

M1 is a simplistic channel model which accounts only for deterministic
path loss as determined by the Friirs equation (see Table~\ref{tab:chanmod}).
This model assumes an exponential decay of received power as a function
of distance, $d$, with an exponent $\eta=2$, i.e., $P_{rx}(d)=K_{F}P_{tx}/d^{2}$,
where $K_{F}$ is a suitable constant.
\begin{table}
\caption{Selected channel models}\vspace*{-2mm}
\label{tab:chanmod}\includegraphics[width=1\columnwidth]{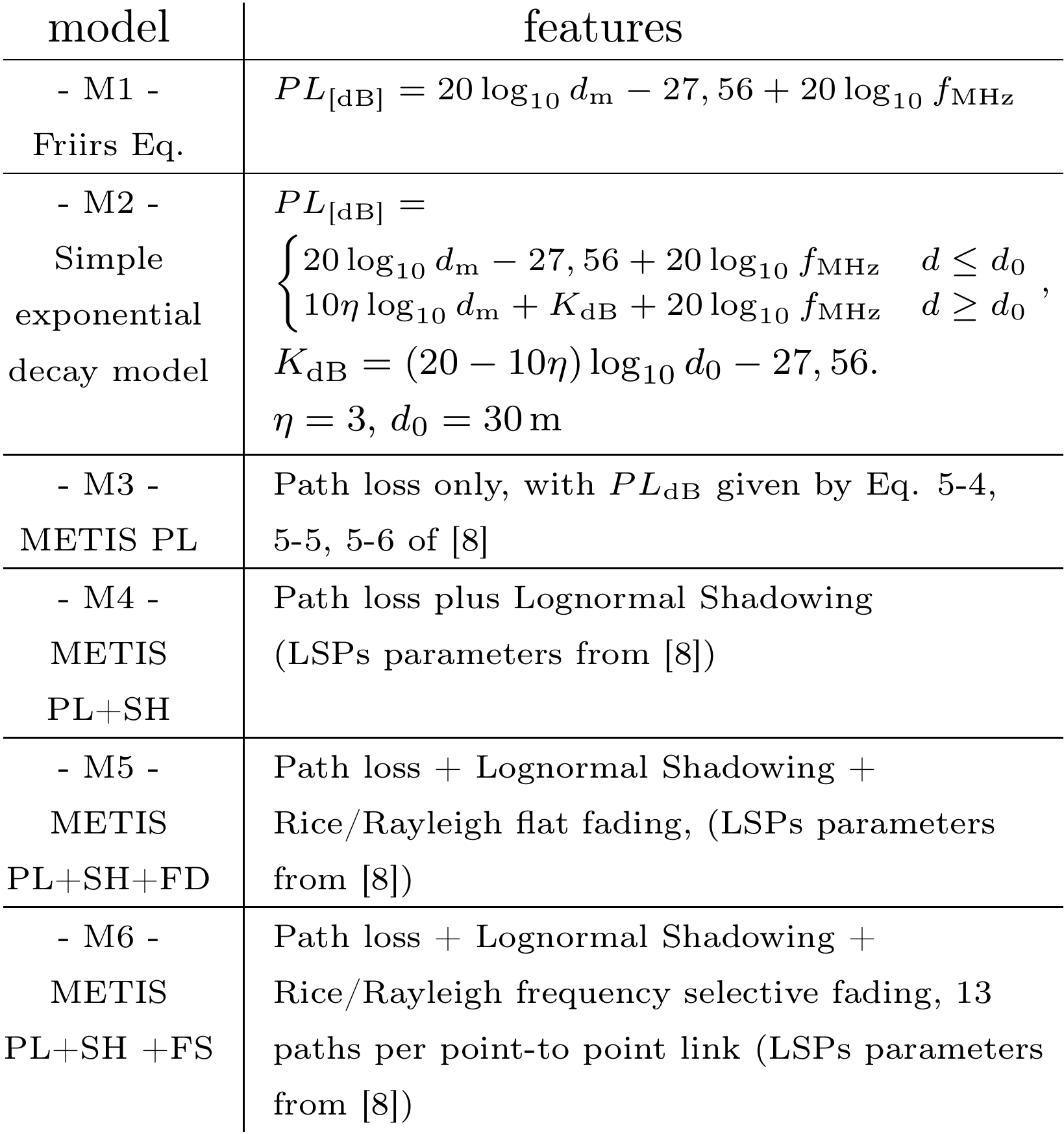}\vspace*{-8mm}
\end{table}

M2 is a similar exponential decay model, with path loss exponent\footnote{In order to be consistent with realistic power attenuation values,
this model is typically considered valid for distances above a close-in
reference distance $d_{0}$. For distances below $d_{0}$, a physical
model like the Friirs equation is typically used. The selection of
$d_{0}$, besides that of the path loss exponent, has a \emph{considerable
}effect on the overall path loss\emph{.}} $\eta=3$.

M3 PL is still a deterministic path loss model, which uses the path
loss formulas 5-4, 5-5, and 5-6 of \cite{METIS_chanmod}.

M4 - PL+SH, includes path loss and \emph{spatially correlated }Lognormal
shadowing, according to the parameters provided by \cite{METIS_chanmod}.

M5 - PL+SH+FD, adds Rayleigh/Rician small scale fading (assumed to
be flat across the frequency domain) on top of shadowing, with spatially
correlated Rician K factors.

M6 - PL+SH+FS, includes frequency selective Rayleigh/Rician small
scale fading. Using M6 entails the generation of the channel transfer
function, evaluated at each subcarrier. The procedure to generate the transfer function taking in input the
LSPs and other parameters, like number of scattering elements for
each channel, delay spread proportionality factor, etc..., is provided in \cite{METIS_chanmod}.

We assume omnidirectional antennas and radiation patterns, and Line
Of Sight (LOS) conditions in each link. Therefore, Rician small scale
fading is used for models M5 and M6. In this work, to parameterize
the channel models, we consider the Urban Micro cell and Urban Macro
cell scenario (UMi and UMa, respectively) of \cite{ITUM2135} for
I2D communications, and the D2D channel parameters presented in \cite{METIS_chanmod}
for D2D communications, see also Section~\ref{subsec:Considered-scenarios}.

\begin{figure}[t]
\begin{centering}
\includegraphics[width=1\columnwidth]{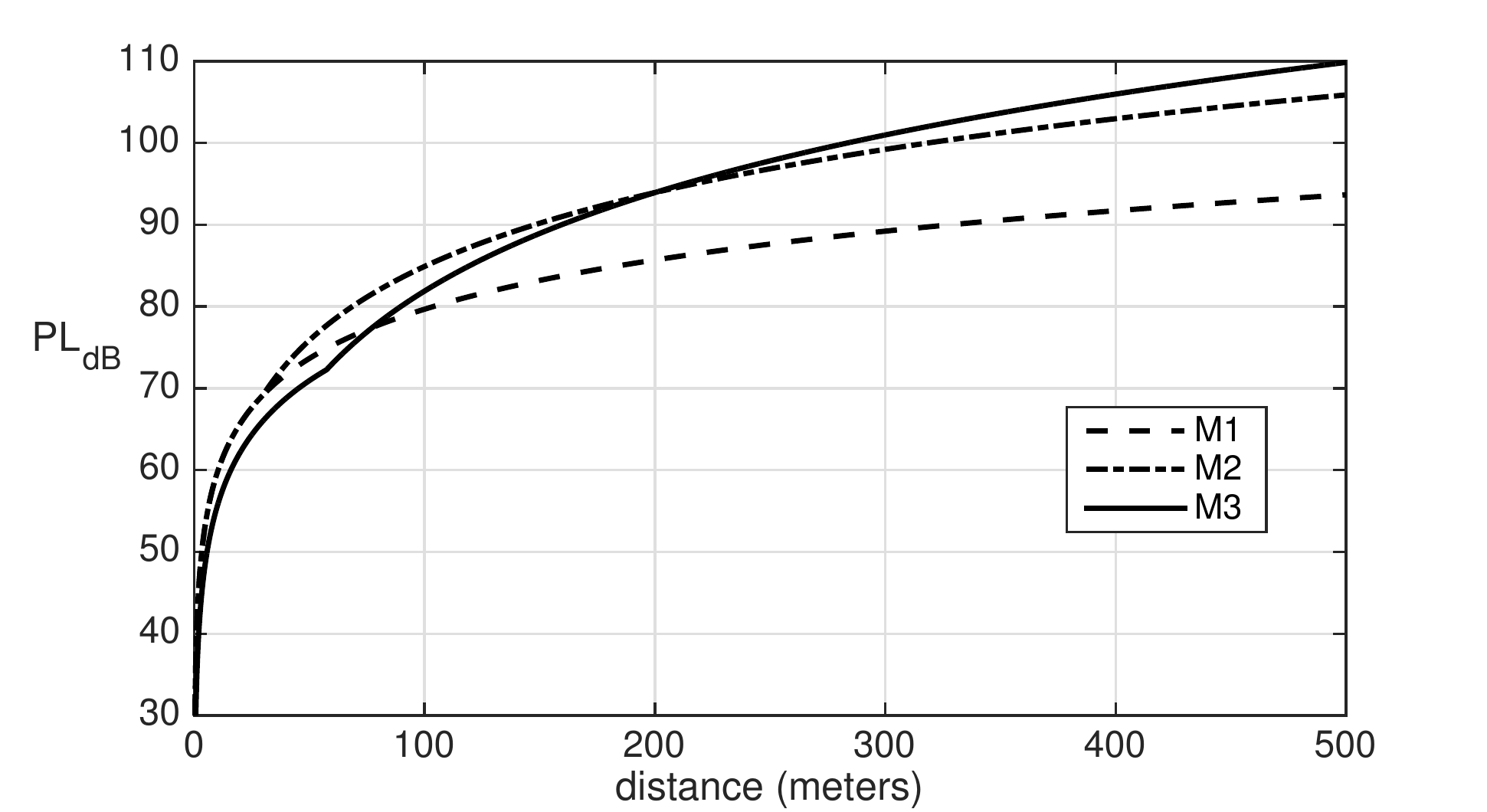}
\par\end{centering}
\caption{Deterministic path loss for models M1-M3.}\label{fig:PL}
\end{figure}

The aim of our study is to evaluate how the performance of the offloading
protocol depend on the channel model in use. To this end, before
presenting our results, it is worth discussing the differences between
the propagation loss entailed by the models M1-M3\footnote{The channel models M4-M6 are obtained by adding, to the propagation
loss of M3, the shadowing (M4), flat fading (M5), and frequency selective
fading (M6), random components. Therefore, Figure~\ref{fig:PL}, and the related
discussion, focus on M1-M3 only.}, and the impact of the different statistics of the random components
for models M4-M6. The following discussion provides useful insights
for the interpretation of the results.

In Figure~\ref{fig:PL}, we plot (in a dB scale) the propagation loss of the
channel models M1-M3 as a function of distance. It can be seen that
both the exponential decay models exhibit deviations from the more
realistic propagation loss model M3 (which has been obtained in \cite{METIS_chanmod}
through large scale measurements campaigns). Particularly, M1 largely
deviates for distances above 200 m, for which it underestimates the
path loss term by more than 10 dB; M2 has the largest deviations (up
to 10 dB), for distances in the range 10-150 m. These deviations may
compromise the reliability of the results of the performance evaluation.
In fact, both M1 and M2 entail an underestimation of the energy spent
(on average) for I2D communications (which can reach distances well
above 200~m), with respect to the energy spent for D2D communications
(which are limited to distances in the order of 100~m). The main
reason is that $PL$, under model M3, has a breakpoint distance (clearly
visible in the figure, at a distance of 60 m), beyond which it increases
(with distance) at a higher rate. Simple exponential decay models,
regardless of the path loss exponent, are inherently smooth, and cannot
reproduce the effect of the breakpoint distance.
\begin{table}[t]
\caption{Selected link margin}
\vspace*{-3mm}
\label{tab:link_margin}\includegraphics[width=1\columnwidth]{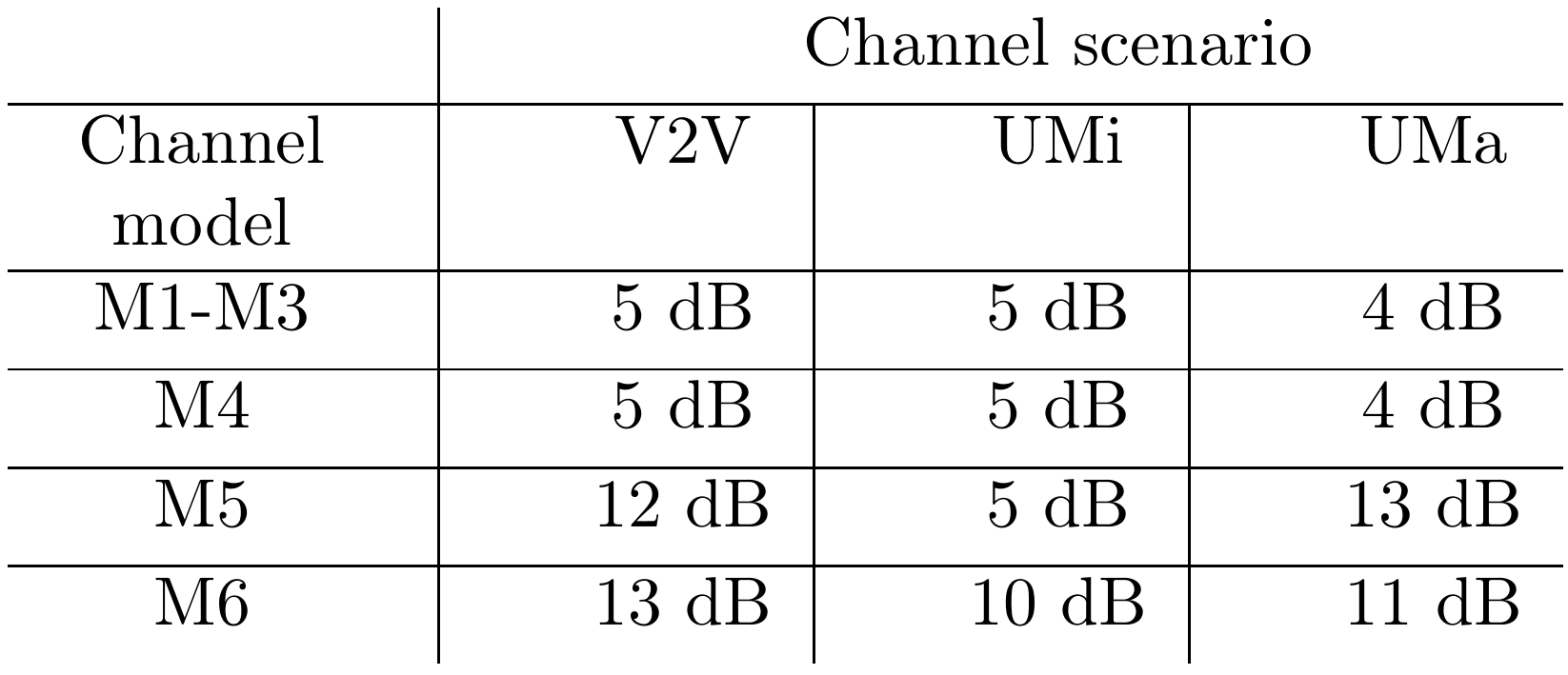}\vspace*{-12mm}
\end{table}

Regarding the random components of the different models, namely Lognormal
shadowing (introduced in M4), flat fading (introduced in M5) and frequency
selective fading (introduced in M6), in Section~\ref{subsec:Forward-error-correction}
we have anticipated that the different combinations of the random
components, result in different statistics of the achievable amount
of information. This affects the link margin selection, which is based
on an outage constraint on the achievable amount of information, making
it dependent on the assumed channel model. In our simulations, we
have set the link margins on the basis of the results of preliminary
montecarlo simulations aimed at determining, for each channel model,
the link margin which guarantees an outage probability below 0,5~\%
assuming the presence of noise only. The results are summarized in
Table~\ref{tab:link_margin}\footnote{For models M1-M3, it is not possible to compute the link margin based
on an outage constraint, because the channel is deterministic, and
hence there is no outage. The link margin, however needs to be present
to guarantee the coexistence among interfering links. Therefore, in
our simulations, we set, for model M1-M3, the same link margin used
for model M4, which, as verified by us in the simulations, allows
the RR scheduler to exploit spatial frequency reuse.}. During our system level simulations (described in the rest of this
section), we have verified that the so selected link margins, thanks
to the effectiveness of the RR scheduler in preventing strongly interfering
links to use the same PRBs, are sufficient to guarantee a below 1\%
outage even in the presence of interfering links.

\subsection{Deployment scenarios and traffic load settings\label{subsec:Considered-scenarios}}

In our simulations, we have considered two deployment scenarios, hereafter
denoted ad A and B, corresponding to a Urban micro-cell scenario and
a Urban macro-cell one, as stadardized by the ITU (see e.g., \cite{ITUM2135}).
In both deployment scenarios the ROI is a street chunk of length 3~Km
and width 20~m. The two scenarios differ in the cell radius (and
number of BSs) and in the antenna height of the BSs, as reported in
Table~\ref{tab:scenario_parameters}.

Vehicles enter the street from both ends according to a Poisson arrival
process with vehicle inter-arrival rate $\lambda_{V}$, and travel
through it at constant speed. The speed of each vehicle is randomly
selected from a uniform distribution in the range $[v_{a},v_{b}]\,\text{m/s}$
and is kept constant during the whole path. We have performed simulations
for three different speed range settings, namely {[}6,16{]}, {[}9,24{]},
and {[}12,32{]} m/s.

The content requests arrival process of each device is modeled as
a Poisson arrival process with inter-arrival rate $\lambda_{C}$,
and the content popularity is modeled by a Zipf distribution
with parameter~$\alpha$, i.e., numbering the available contents
as $k=1,2,\ldots$ , the probability that a device requests content
$k$ is $p(k)=\frac{1}{\zeta(\alpha)}k^{-\alpha}$, where $\zeta(\cdot)$
is the Riemann $\zeta$-function. 
\begin{table}
\begin{centering}
\caption{Deployment scenarios parameters}
\label{tab:scenario_parameters}
\par\end{centering}
\centering{}\includegraphics[width=1\columnwidth]{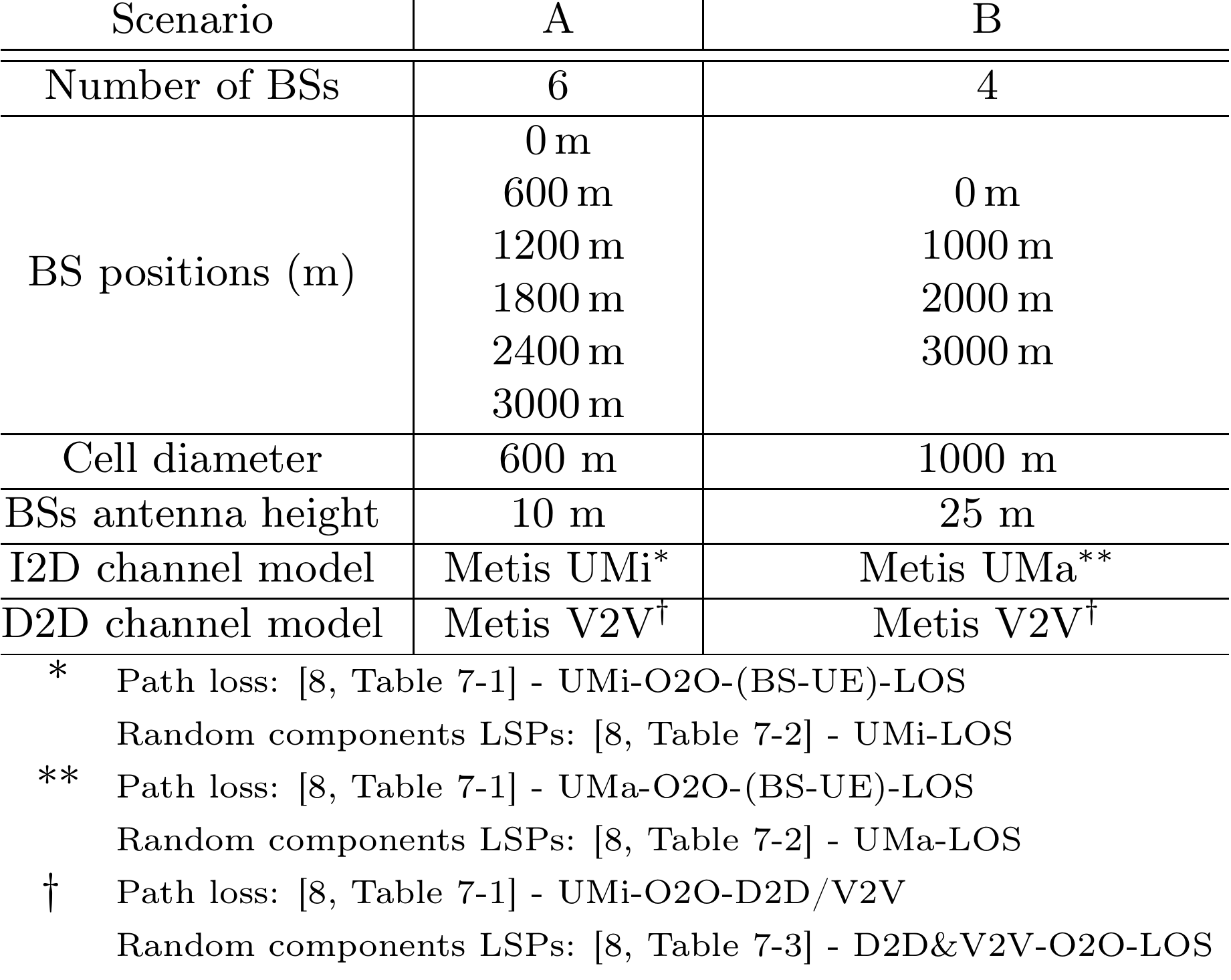}\hrule\vspace*{-6mm}
\end{table}

We set the following parameters which determine the traffic load:
$\lambda_{V}=20$ new vehicles per minutes, each vehicles generates
content requests with an average rate $\lambda_{C}=6$ requests per
minute, and the content payload size is $L=432\,\text{kB}$. 

We measure the average offered traffic load in terms of a traffic
load spatial density $\rho_{Q}$. Since our scenario is (essentially)
unidimensional, we consider a linear density (instead of a surface
density), and express $\rho_{Q}$ in $\text{kbps/m}$. Hence, $\rho_{Q}$
represents the average number of bits requested by the devices located
in a street chunk of length 1~m (and width equal to the street width).
$\rho_{Q}$ is the product of the average number of vehicles present
in a street chunk of length $1\,\text{m}$, say $\rho_{V}$\footnote{For a uniform distribution of the speed range in an interval $[v_{a},v_{b}]$
m/s, the average linear density of the vehicles is given by $\rho_{V}=\lambda_{V}\left(\ln\left(v_{a}^{-1}\right)-\ln\left(v_{b}^{-1}\right)/(v_{b}-v_{a})\right)$
\cite{Pescosolido2018WoWMoM}.}, times the content request arrival rate per user $\lambda_{C}$,
times the coded packet size $D$, times the ratio, which we indicated
with $\gamma_{\text{nr}}$, of non-repeated request to total requests\footnote{Since the content requests are assumed to be i.i.d, there is a non
negligible probability that a device makes the request of a content
it already has in its cache. Repeated requests are simply discarded
by the CDMS, whereas the sharing timeout counter is reset to its starting
value $\tau_{s}$. In \cite[Lemma 4]{Pescosolido2018WoWMoM}, we computed
the ratio of non-repeated to total requests, here indicated with $\gamma_{\text{nr}}$.
For the specific settings considered here, for content request processes,
we obtain $\gamma_{\text{nr}}\simeq0.59$.}, i.e., $\rho_{Q}=\lambda_{C}\rho_{V}D$$\gamma_{\text{nr}}$ $\text{kbps/m}$.
For the three considered vehicles speed ranges {[}6,16{]}, {[}9,24{]},
and {[}12,32{]} m/s, we obtain $\rho_{Q}=13.9$, $9.26$, and $6.95\,\text{kbps/m}$,
respectively. 

A conventional cellular system with BSs deployed on a straight line,
a spatial frequency reuse factor of 1/3, and cell diameter $d_{c}$,
can support a maximum traffic load density equal to $\rho_{Q,\max}^{(\text{I2D})}=We/3/d_{c}=36\,\text{\text{kbps/m}}$,
considering the cell diameter of Scenario A, and $21\,\text{kbps/m}$,
considering scenario B.

All the system parameters are summarized in Table~\ref{tab:sysparam}.\vspace*{-2mm}

\subsection{Simulation results\label{subsec:Simulation-results}}

For each deployments scenario (A,B), each traffic model input, and
each channel model M1-M6, we run 10 simulations using i.i.d. random
realizations of the vehicles arrival process, the content requests
process, and (for channel models M4-M6 only, since M1-M3 are deterministic
models) the spatially correlated random channels set. For each realization
of the random temporal and spatial processes, we run a system simulation
of 30 minutes\footnote{To enable a fair comparison among the channel models, we used, for
each channel model, the same 10 i.i.d. realizations of the vehicles
arrival and content request processes.}.

\begin{figure}[!t]
\subfloat[Scenario A]{\centering{}\includegraphics[width=0.9\columnwidth]{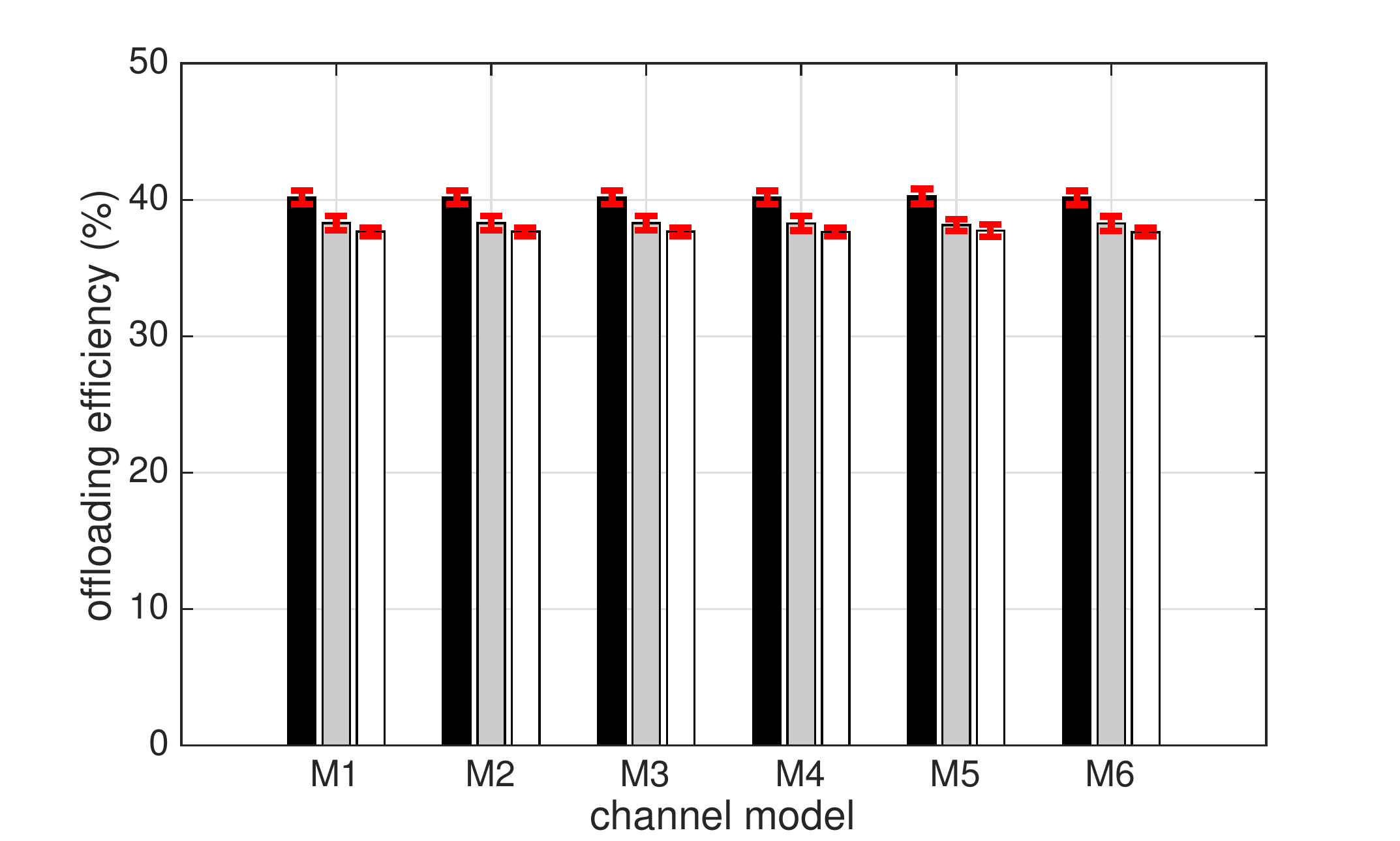}}
\vspace*{-3mm}
\subfloat[Scenario B]{\begin{centering}
\includegraphics[width=0.9\columnwidth]{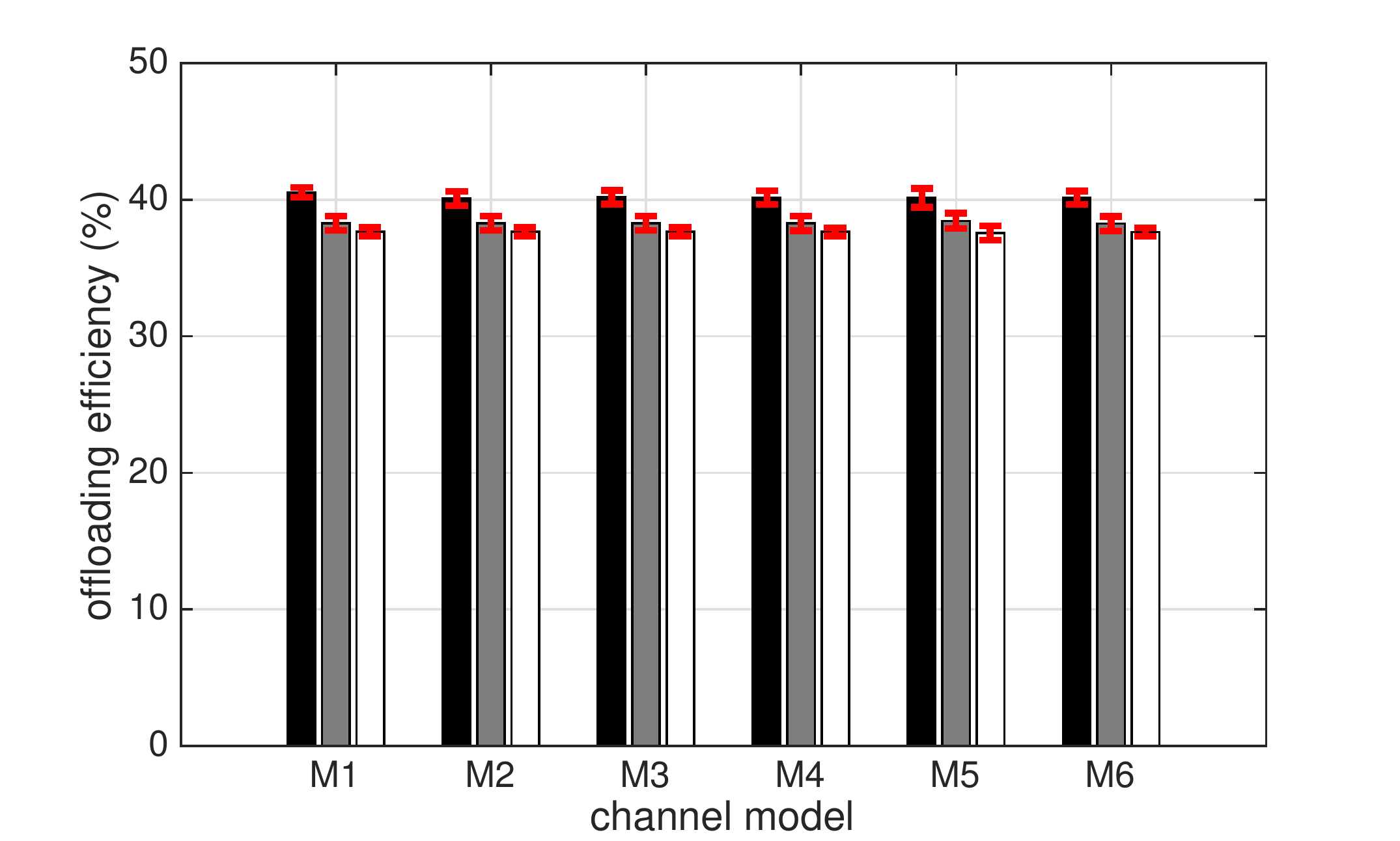}
\par\end{centering}
}\caption{Offloading efficiency in deployment scenario A and B, for speed range
{[}6,16{]} (black bars), {[}9,24{]} (grey bars), {[}12,32{]} (white
bars).}\label{fig:R1_OFF_EFF}
\vspace*{-7mm}
\end{figure}

Figure~\ref{fig:R1_OFF_EFF} displays the offloading efficiency (with 95\% confidence
intervals), defined as the percentage of contents delivered through
offloading, i.e., using D2D communications, achieved under scenarios
A (Figure~\ref{fig:R1_OFF_EFF}.a) and B (Figure~\ref{fig:R1_OFF_EFF}.b). Each bar color refers to a different
speed range setting. In general, with the considered vehicles traffic
parameters, an offloading efficiency around 40\% is achieved. It can
be seen that, with higher speed ranges, there is a decrease of the
offloading efficiency due to the decreasing vehicles density. The
decrease of offloading efficiency, however, is quite small. 
\begin{table}[t]
\caption{System parameters used for performance evaluation}
\vspace*{-3mm}
\label{tab:sysparam}\includegraphics[width=1\columnwidth]{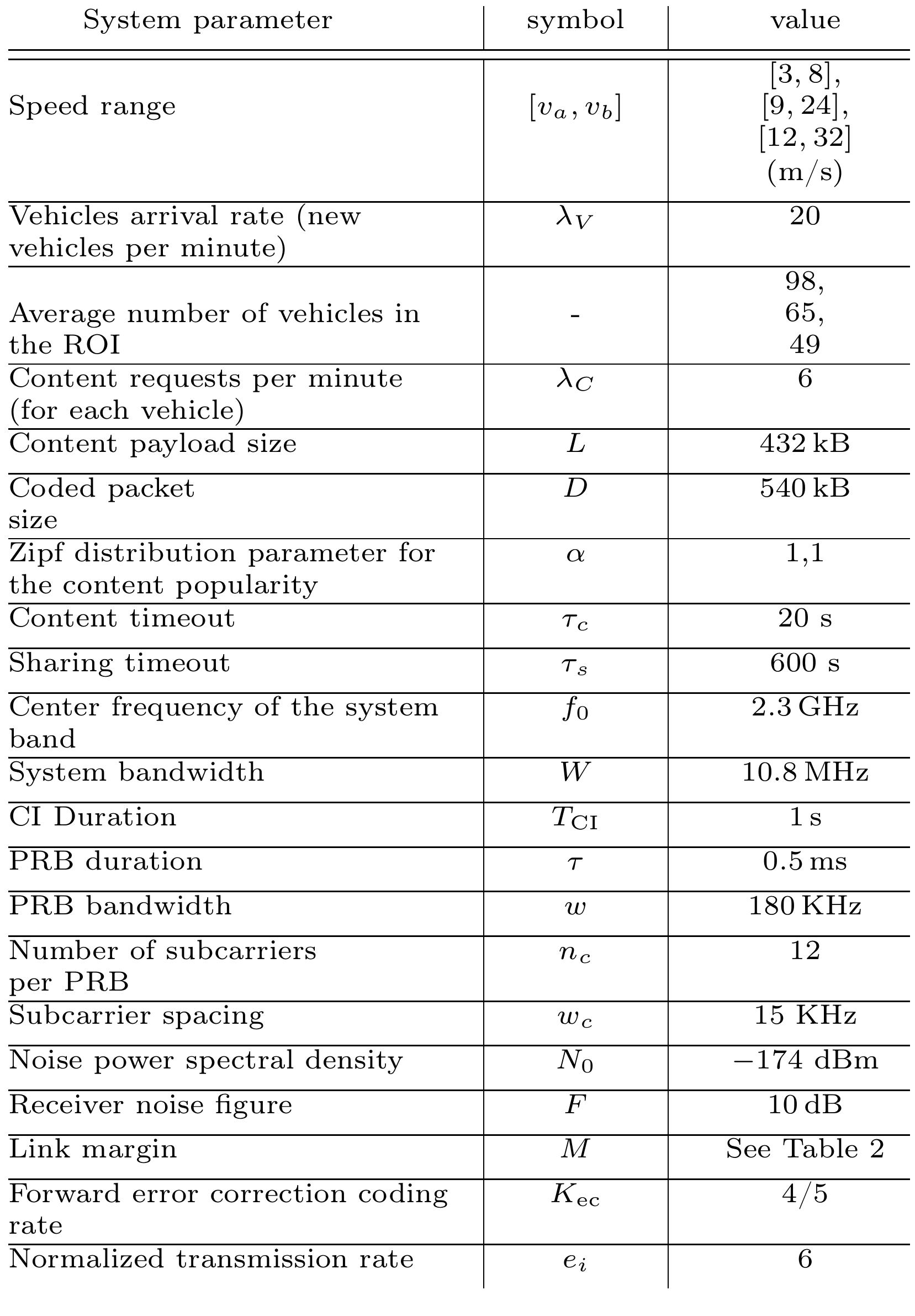}\vspace*{-10mm}
\end{table}
This is the effect of delay-tolerance: with higher speed ranges, the
number of vehicles encountered by each device during the content timeout
of each request increases, thus counterbalancing the decreased vehicles
density. In the absence of delay tolerance, the decrease in offloading
efficiency would be much higher. Offloading efficiency is not affected
either by the deployment scenario, or the channel model. This is due
to the fact that the CDMS schedules D2D offloading as a function of
nominal channel gain rankings, which depend only on distance.

\begin{figure*}[!t]
\subfloat[Benchmark I2D-only scheme]{\centering{}\includegraphics[width=0.9\columnwidth]{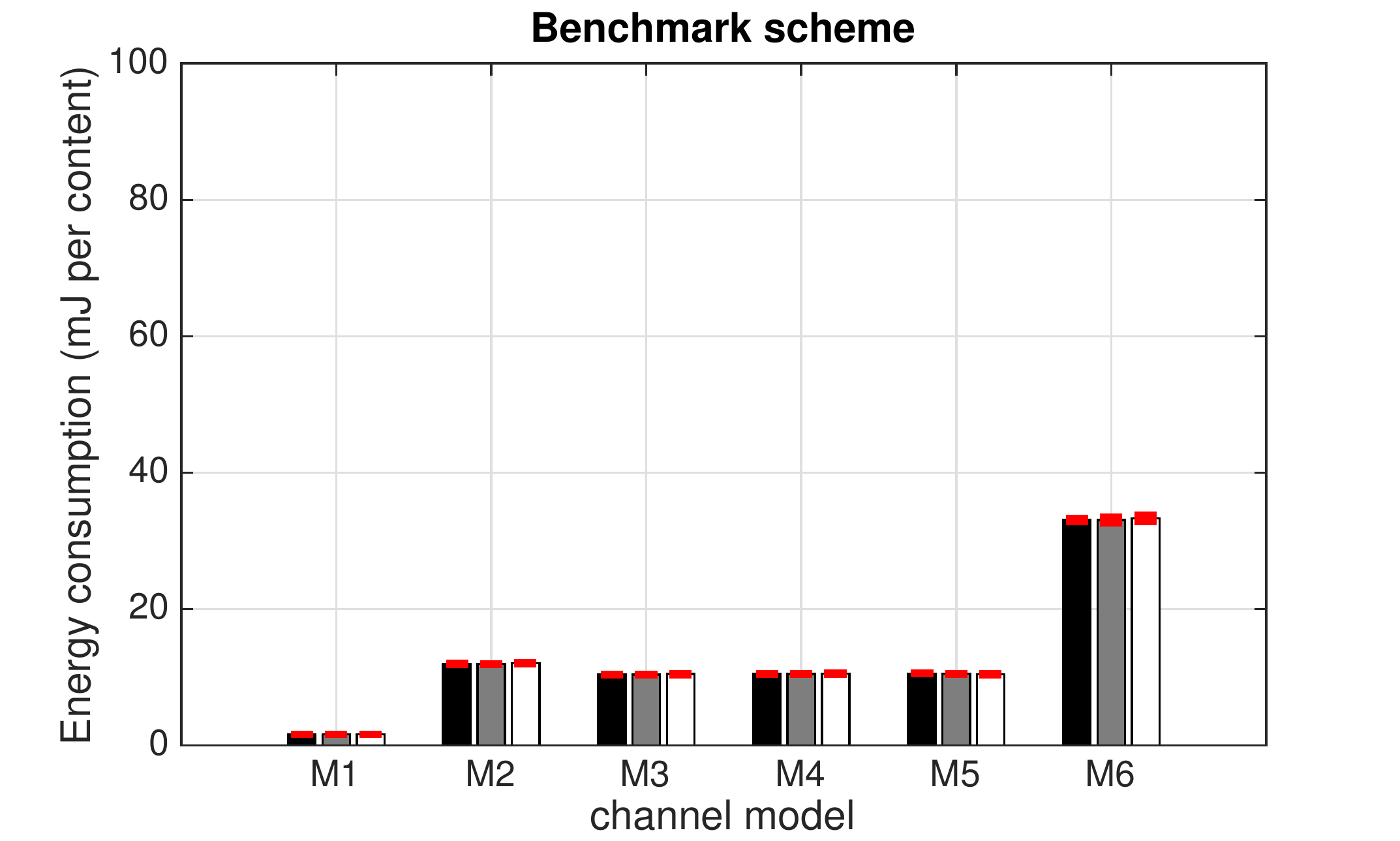}}\hfill{}\subfloat[CDMS-aided offloading scheme]{\begin{centering}
\includegraphics[width=0.9\columnwidth]{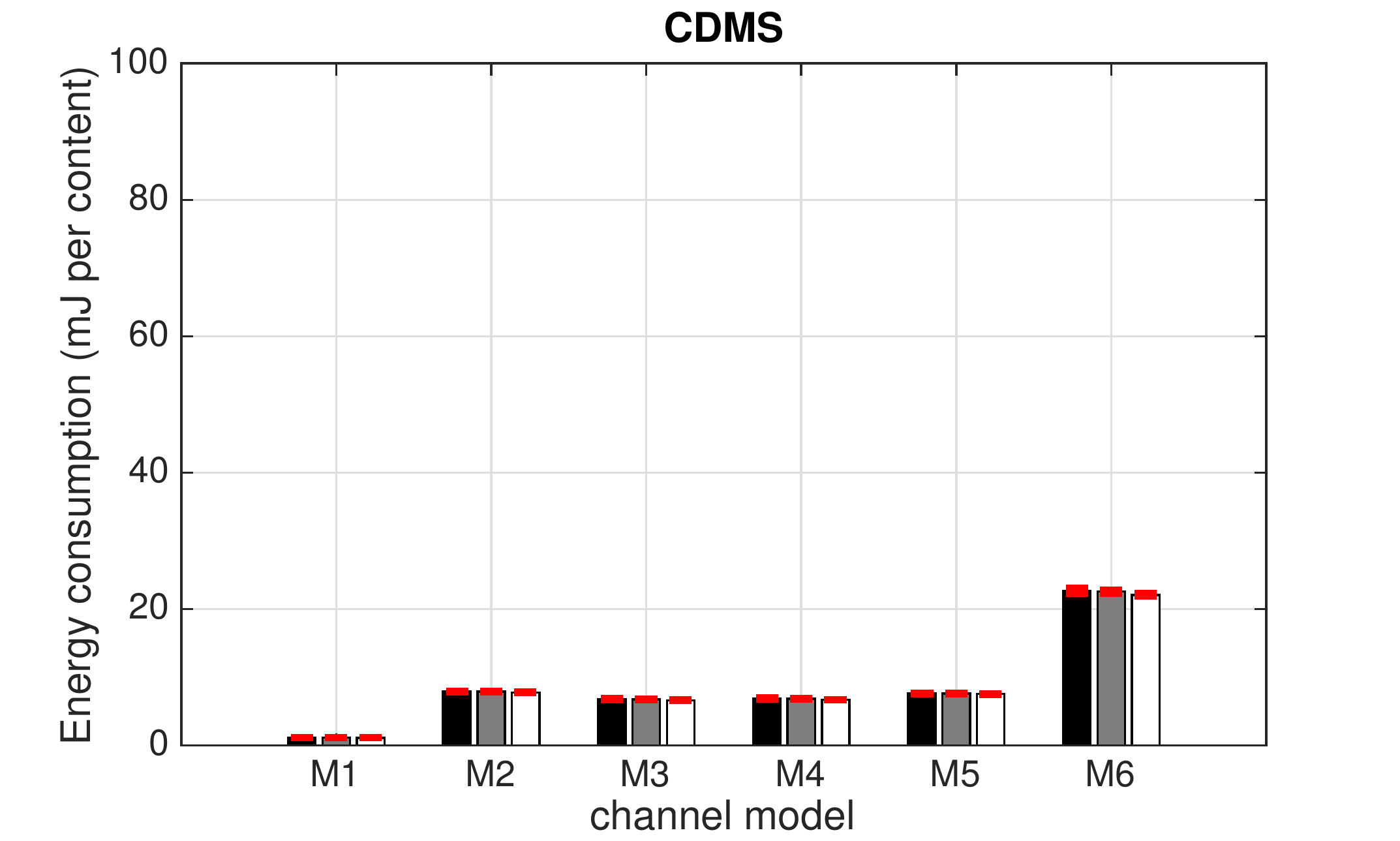}
\par\end{centering}
}\caption{Energy consumption for the benchmark and CDMS-aided offloading schemes
in Deployment Scenario A. Speed range {[}6,16{]} (black bars), {[}9,24{]}
(grey bars), {[}12,32{]} (white bars).}\label{R1_ENERGY_SCEN_A}
\vspace*{-5mm}
\end{figure*}

Figure~\ref{R1_ENERGY_SCEN_A} displays the average energy consumption (with $95\%$ confidence
intervals over the set of simulations) of the entire system (i.e.,
including the energy spent by both I2D and D2D content deliveries)
measured in mJ per delivered content, obtained in Scenario~A. Figure~\ref{R1_ENERGY_SCEN_A}.a
refers to the benchmark I2D-only scheme, and Figure~\ref{R1_ENERGY_SCEN_A}.b to the CDMS-aided
D2D data offloading scheme.

\begin{figure*}[!t]
\subfloat[Benchmark I2D-only scheme]{\centering{}\includegraphics[width=0.9\columnwidth]{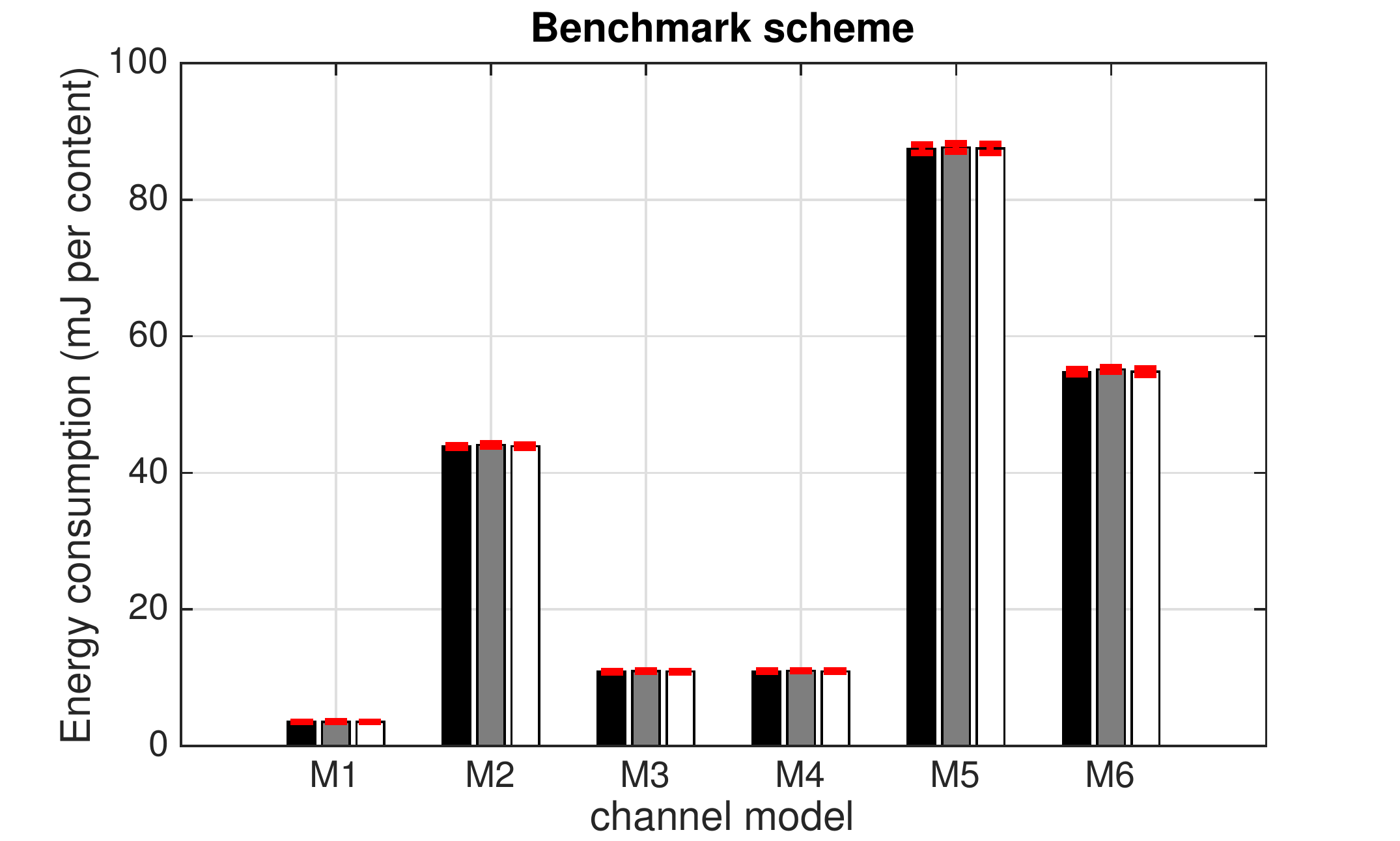}}\hfill{}\subfloat[CDMS-aided offloading scheme]{\begin{centering}
\includegraphics[width=0.9\columnwidth]{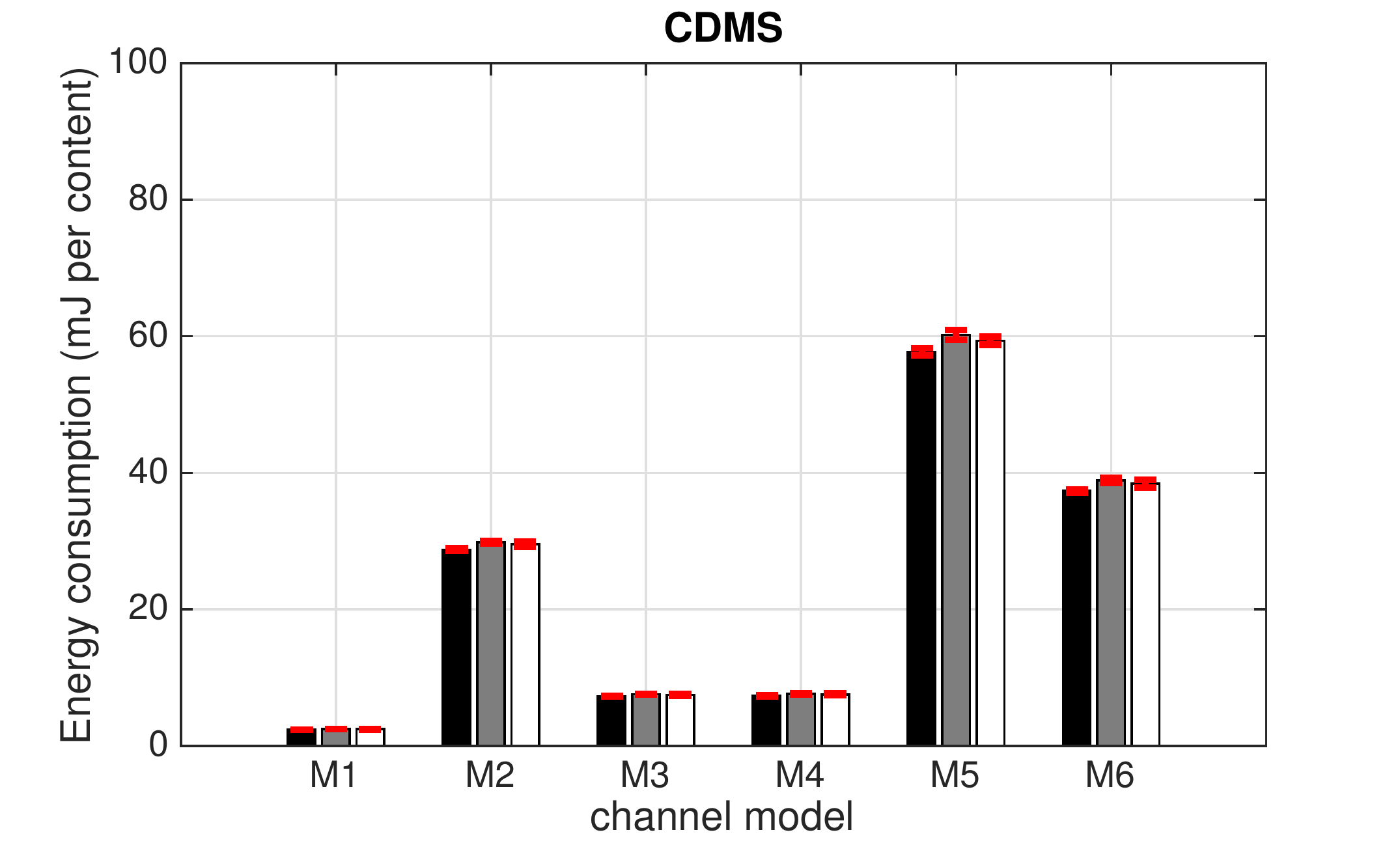}
\par\end{centering}
}\caption{Energy consumption for the benchmark and CDMS-aided offloading schemes
in Deployment Scenario B. Speed range {[}6,16{]} (black bars), {[}9,24{]}
(grey bars), {[}12,32{]} (white bars).}\label{R1_ENERGY_SCEN_B}
\end{figure*}

Figure~\ref{R1_ENERGY_SCEN_B} refers to the same performance metric, obtained under Scenario~B.
It can be seen that there is a considerable mismatch among the performance
obtained using different channel models. Notably, the flat fading
model M5, in the case of Scenario B, tends to overestimate the energy
consumption, with respect to the frequency selective model M6, whereas
in Scenario A it is the opposite. This is due to the fact that the
required power link margin for I2D communications, computed using
M5, is higher than the one computed using M6 under Scenario B, and
lower under Scenario A (see Table 2). The link margin, computed on
the basis of an outage constraint, captures specific differences in
the statistics of the frequency selective channels behavior macro-cells
and micro-cells. In general, higher antenna heights (25 m under Scenario
B, opposed to 10 m under Scenario A) entail a lighter tail of probability
density function of the achievable amount of information, which turns
out in lower required link margin to satisfy the outage constraint.
This realistic behavior, however, is not well captured by the flat
fading channel model M5, thus leading to the mismatch.

\begin{figure*}[!t]
\subfloat[Scenario A]{\centering{}\includegraphics[width=0.9\columnwidth]{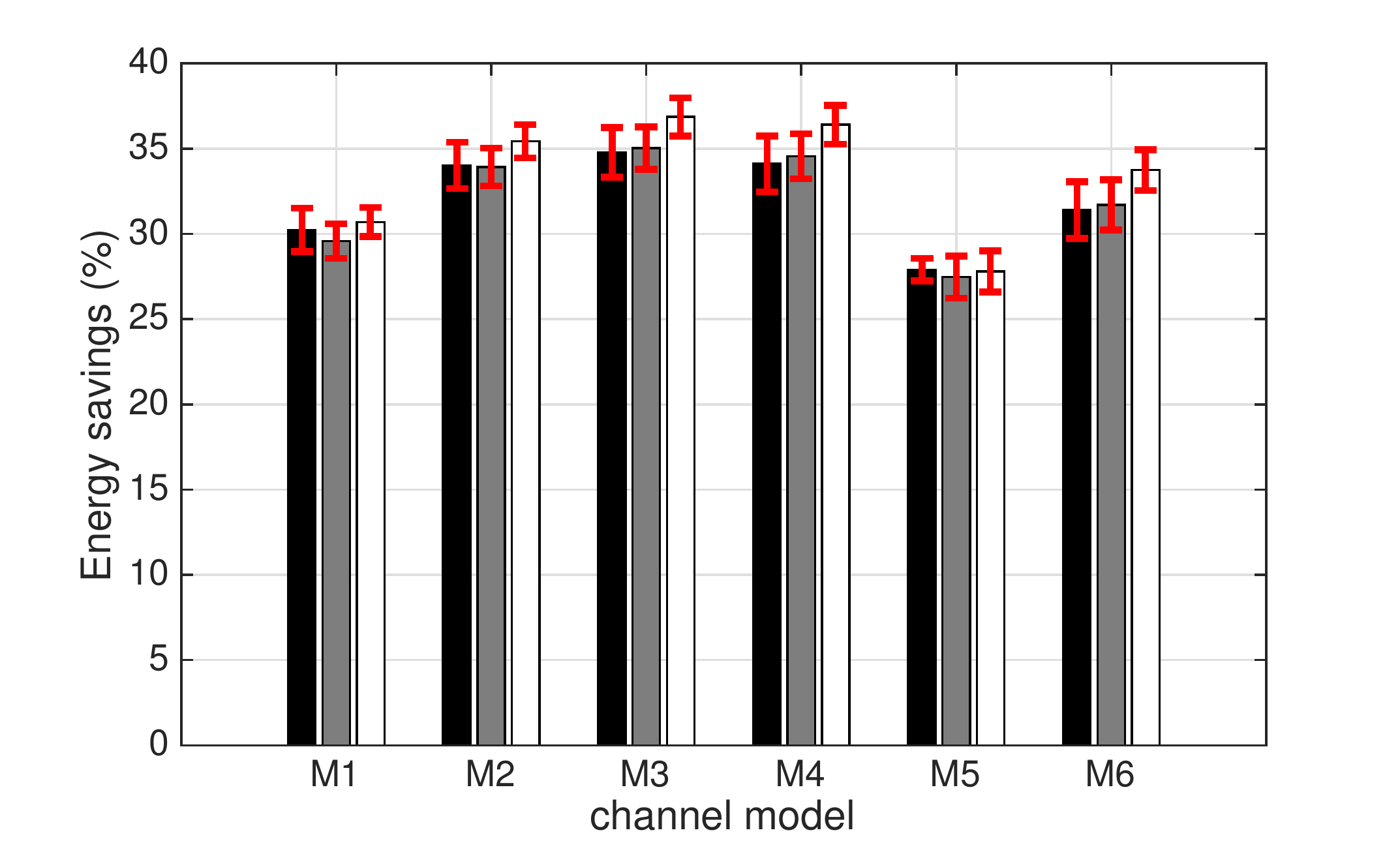}}\hfill{}\subfloat[Scenario B]{\begin{centering}
\includegraphics[width=0.9\columnwidth]{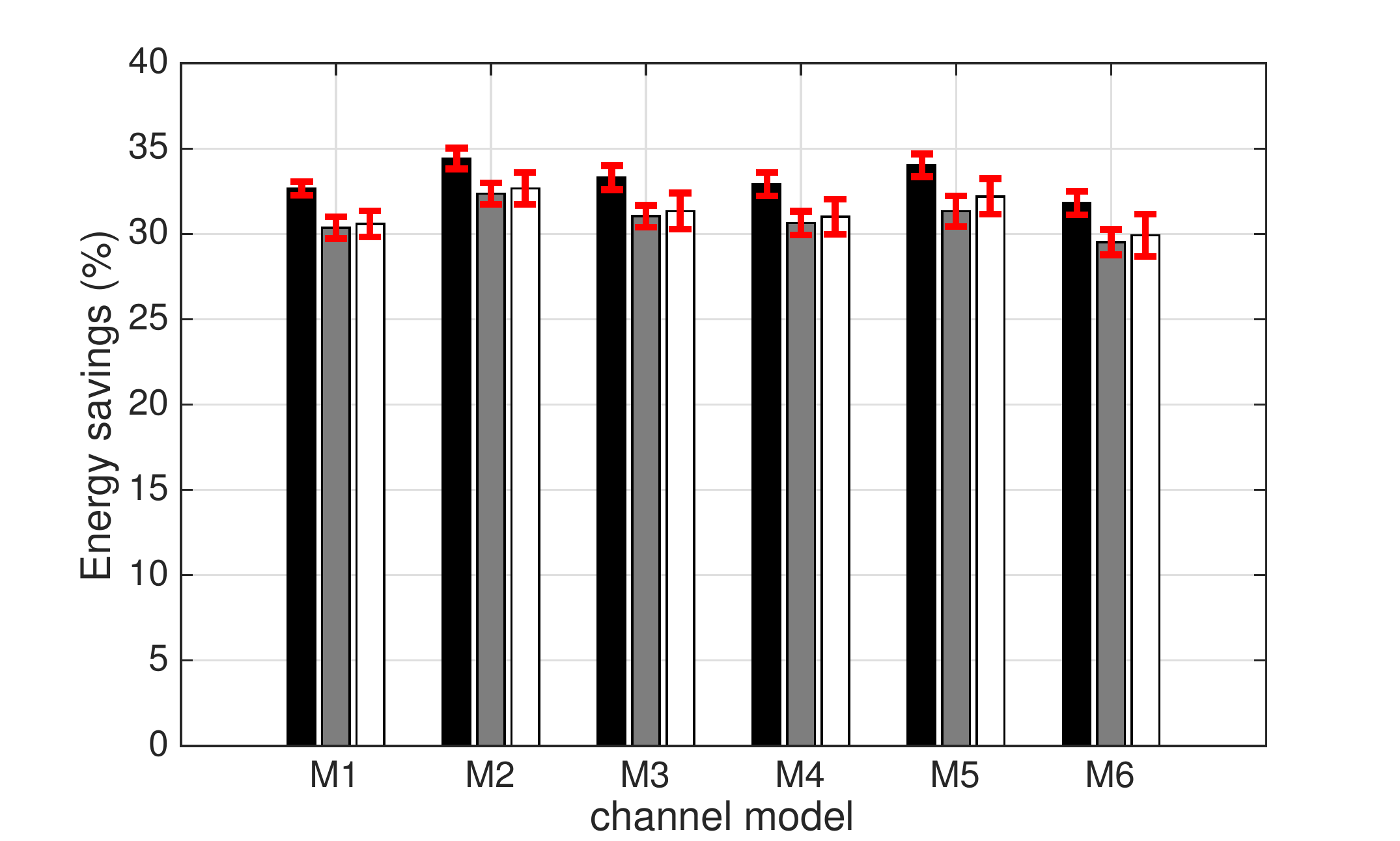}
\par\end{centering}
}\caption{Energy savings percentage in Scenarios A and B; speed range {[}6,16{]}
(black bars), {[}9,24{]} (grey bars), {[}12,32{]} (white bars).}\label{R1_ENERGY_SAVINGS}
\vspace*{-4mm}
\end{figure*}

Figure~\ref{R1_ENERGY_SAVINGS} shows the estimation of the energy savings percentage of
the CDMS-aided D2D offloading scheme with respect to the benchmark
scheme under Scenario A (Figure~\ref{R1_ENERGY_SAVINGS}.a) and Scenario B (Figure~\ref{R1_ENERGY_SAVINGS}.b).
With Scenario B, the estimation seems to be consistent for all the
channel models (with deviation below 2\%). Under Scenario A, however,
mismatches in the order of 4\% energy saving can be observed, with
the worst performing model M5 entailing a 6\% underestimation
of the achievable energy saving gain.

\begin{figure*}[t]
\subfloat[Benchmark I2D-only scheme]{\centering{}\includegraphics[width=0.9\columnwidth]{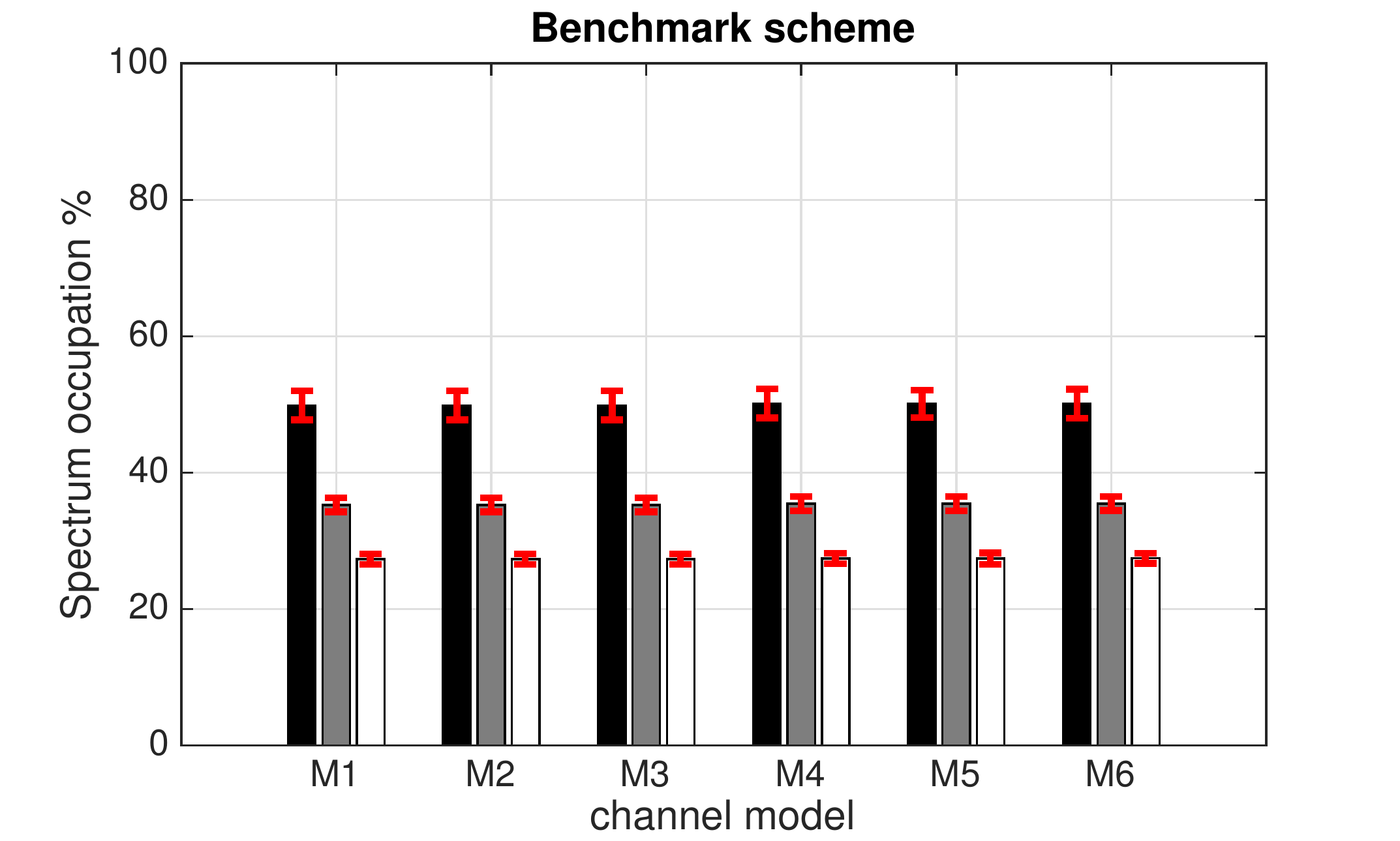}}\hfill{}\subfloat[CDMS-aided offloading scheme]{\begin{centering}
\includegraphics[width=0.9\columnwidth]{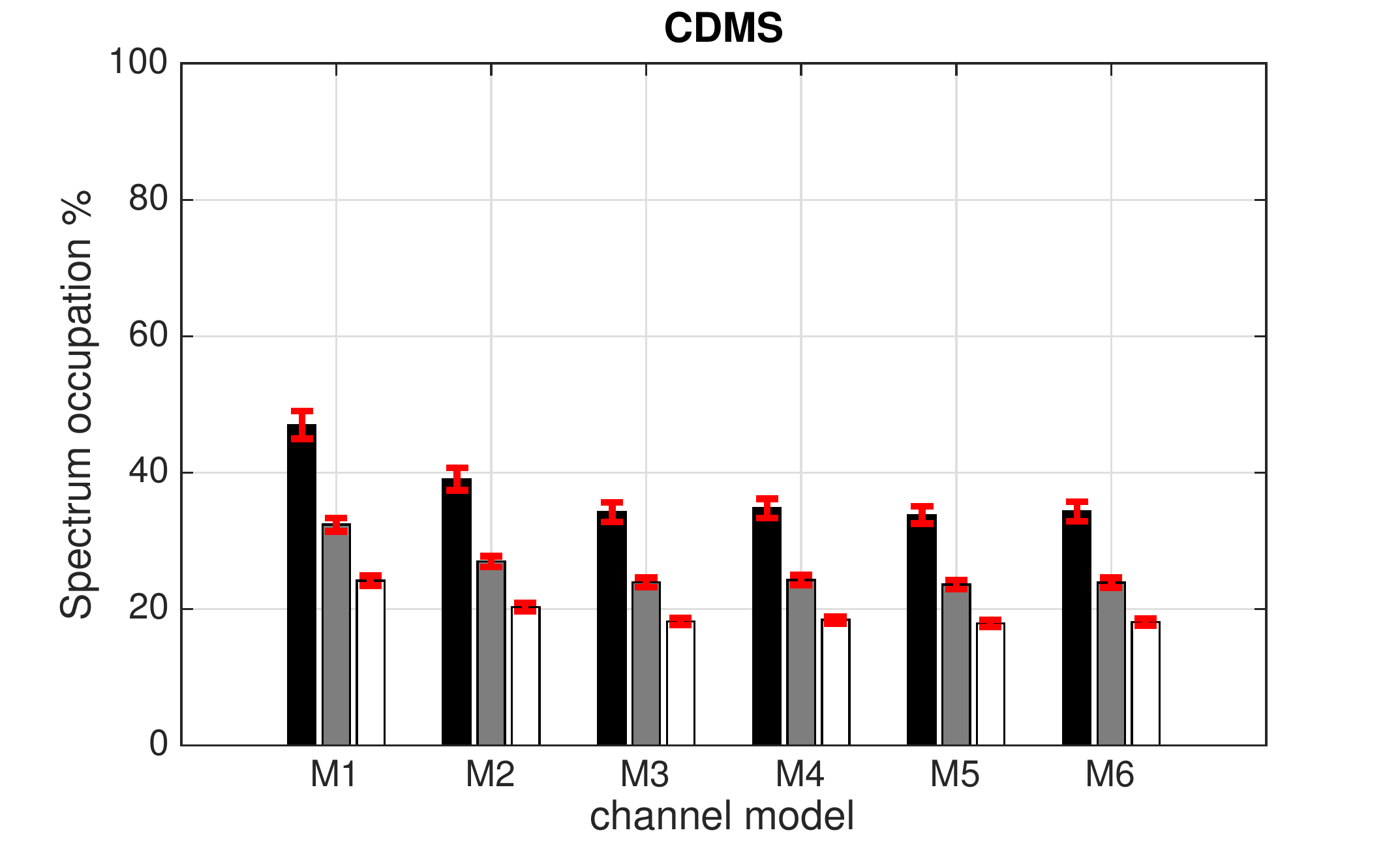}
\par\end{centering}
}\caption{Spectrum occupation percentage obtained with deployment scenario A
for the benchmark and the CDMS-aided schemes; speed range {[}6,16{]}
(black bars), {[}9,24{]} (grey bars), {[}12,32{]} (white bars).}\label{R1_SPECTRUM_A}
\vspace*{-2mm}
\end{figure*}

\begin{figure*}[!t]
\subfloat[Benchmark I2D-only scheme]{\centering{}\includegraphics[width=0.9\columnwidth]{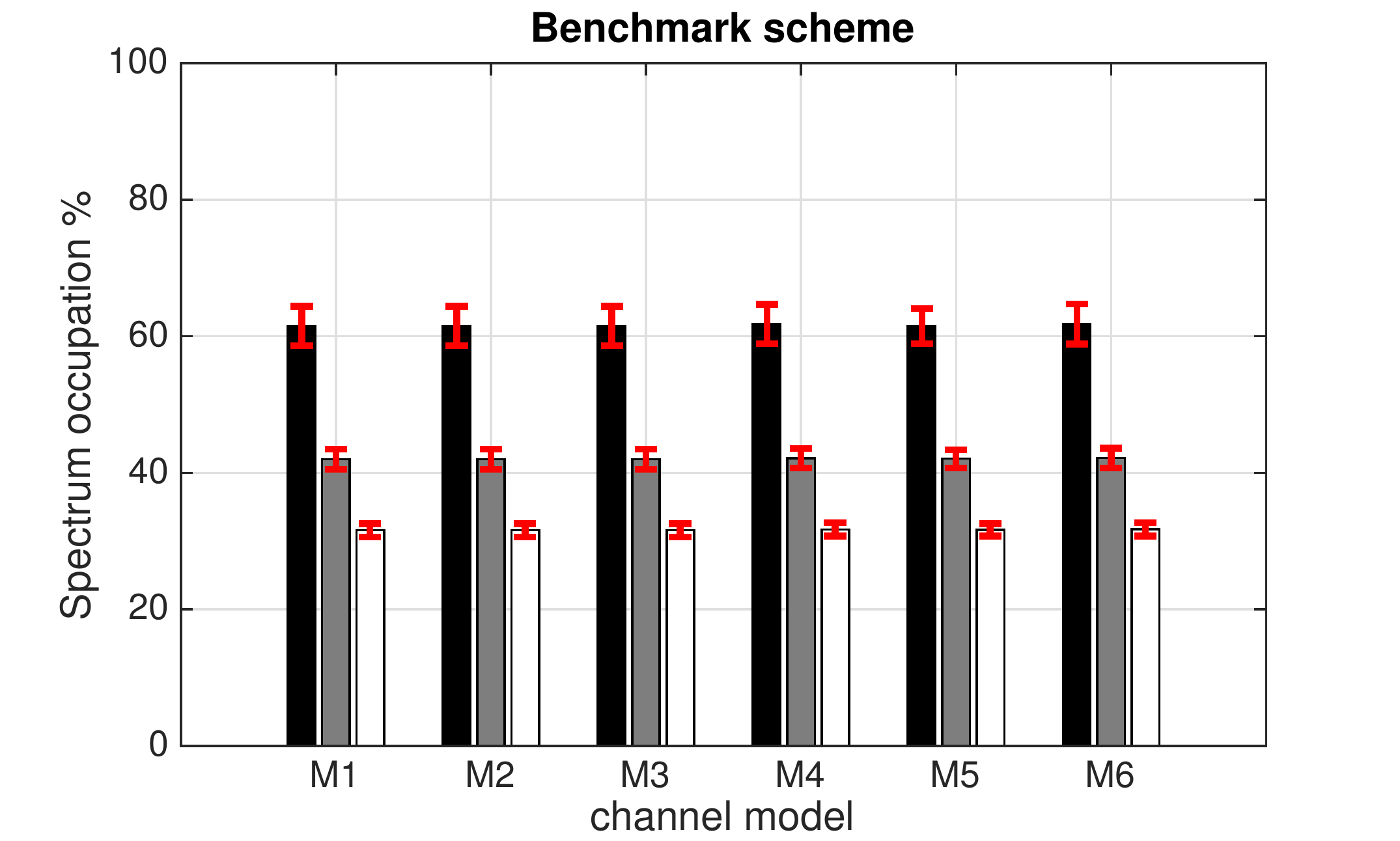}}\hfill{}\subfloat[CDMS-aided offloading scheme]{\begin{centering}
\includegraphics[width=0.9\columnwidth]{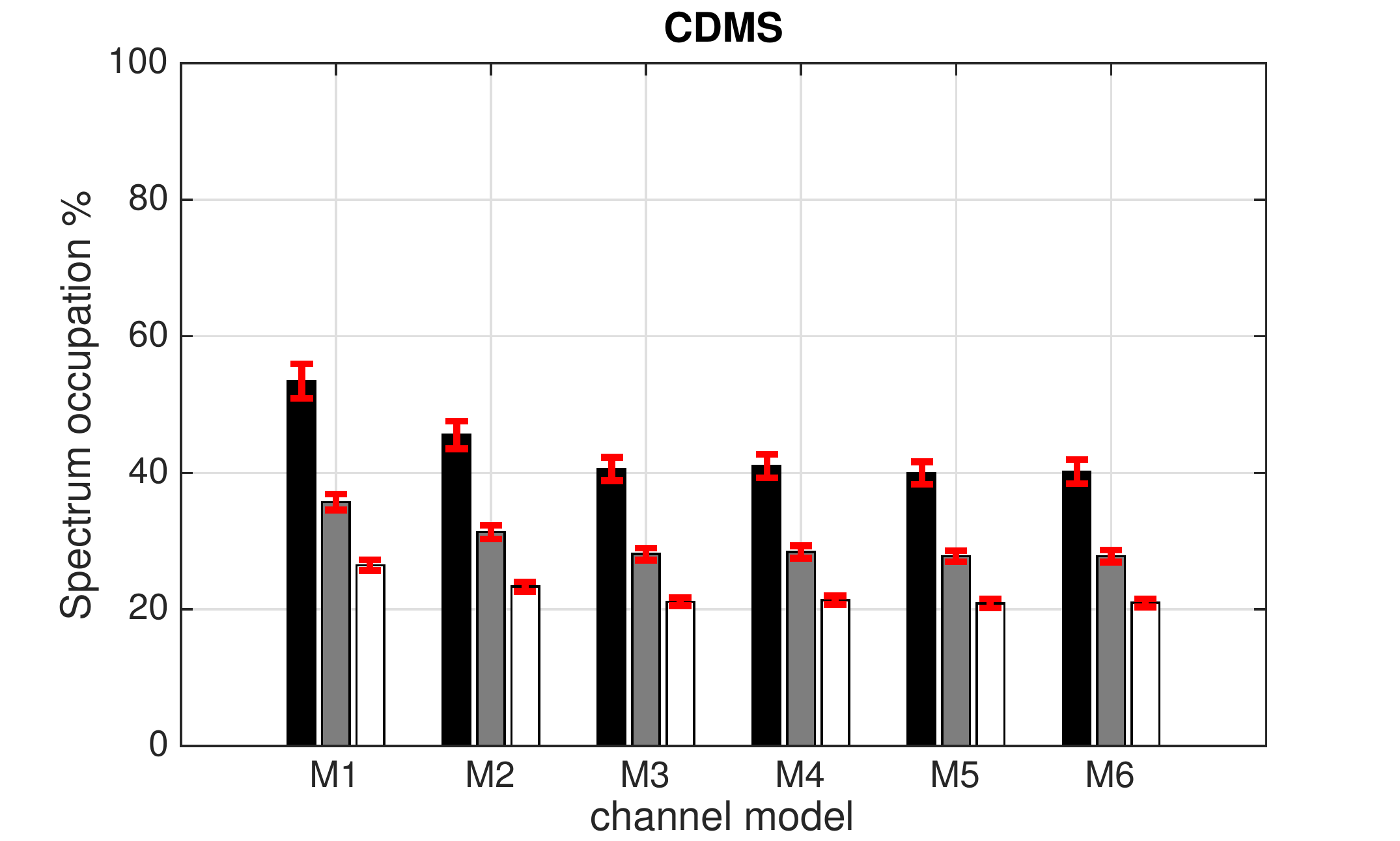}
\par\end{centering}
}\caption{Spectrum occupation percentage obtained with deployment scenario B
for the benchmark and the CDMS-aided schemes; speed range {[}6,16{]}
(black bars), {[}9,24{]} (grey bars), {[}12,32{]} (white bars).}\label{R1_SPECTRUM_B}
\vspace*{-4mm}
\end{figure*}

Figures~\ref{R1_SPECTRUM_A}~and~\ref{R1_SPECTRUM_B} report the percentage of the spectrum usage for scenarios
A and B, respectively. In each CI, each PRB is counted as being used
if it is allocated to at least one communication over the entire ROI.
The effect of spatial frequency reuse is visible since the CDMS-aided
offloading scheme (Figures~\ref{R1_SPECTRUM_A}.b and~\ref{R1_SPECTRUM_B}.b) uses less PRBs than the benchmark
scheme (Figures~\ref{R1_SPECTRUM_A}.a and~\ref{R1_SPECTRUM_B}.a). In this case, the deterministic channel
models M1 and M2 provide unreliable results, overestimating spectrum occupation when considering the CDMS-aided scheme, whereas the results obtained
with models M3-M6 are consistent.

\begin{figure*}[!t]
\subfloat[Scenario A]{\centering{}\includegraphics[width=0.9\columnwidth]{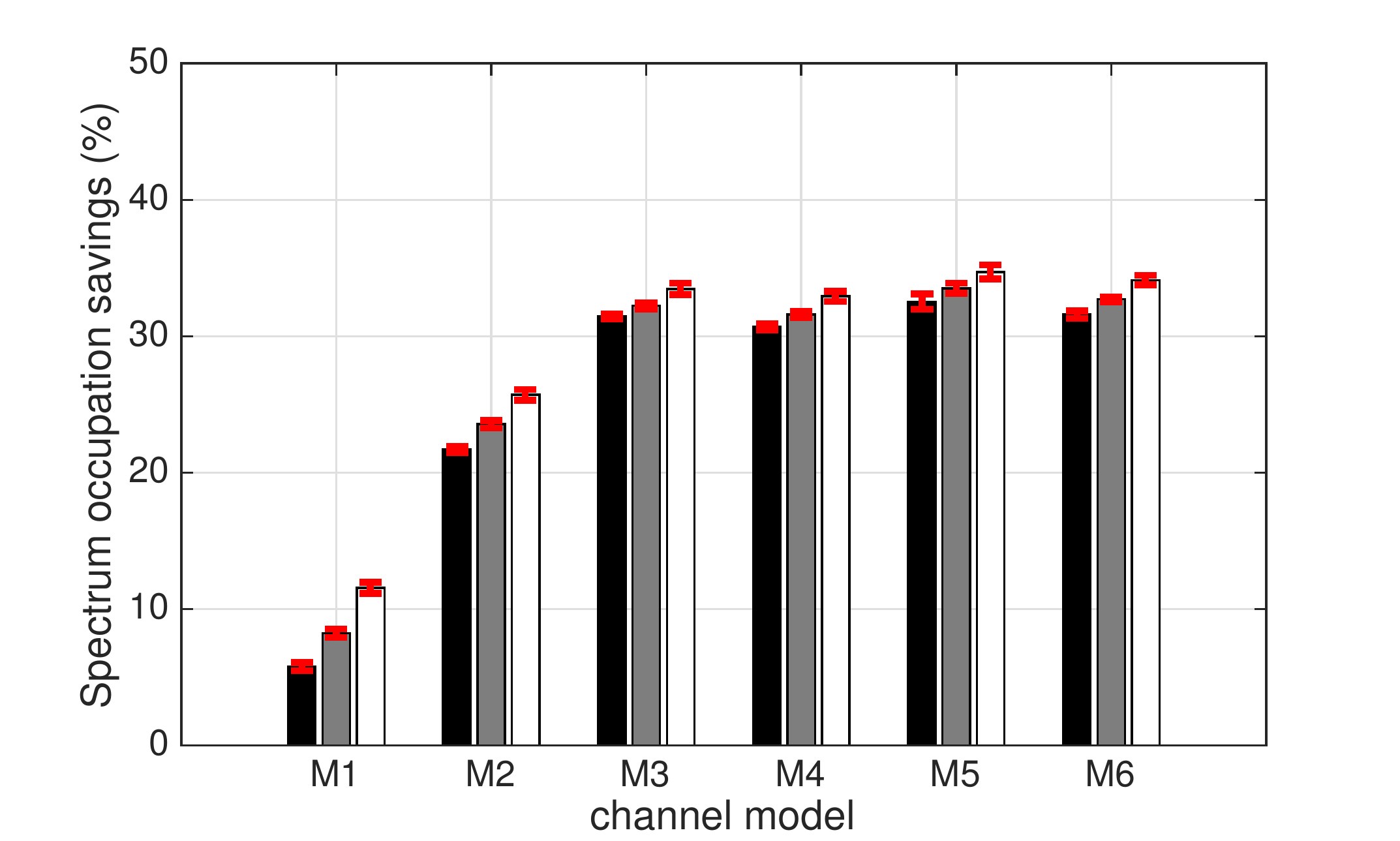}}\hfill{}\subfloat[Scenario B]{\begin{centering}
\includegraphics[width=0.9\columnwidth]{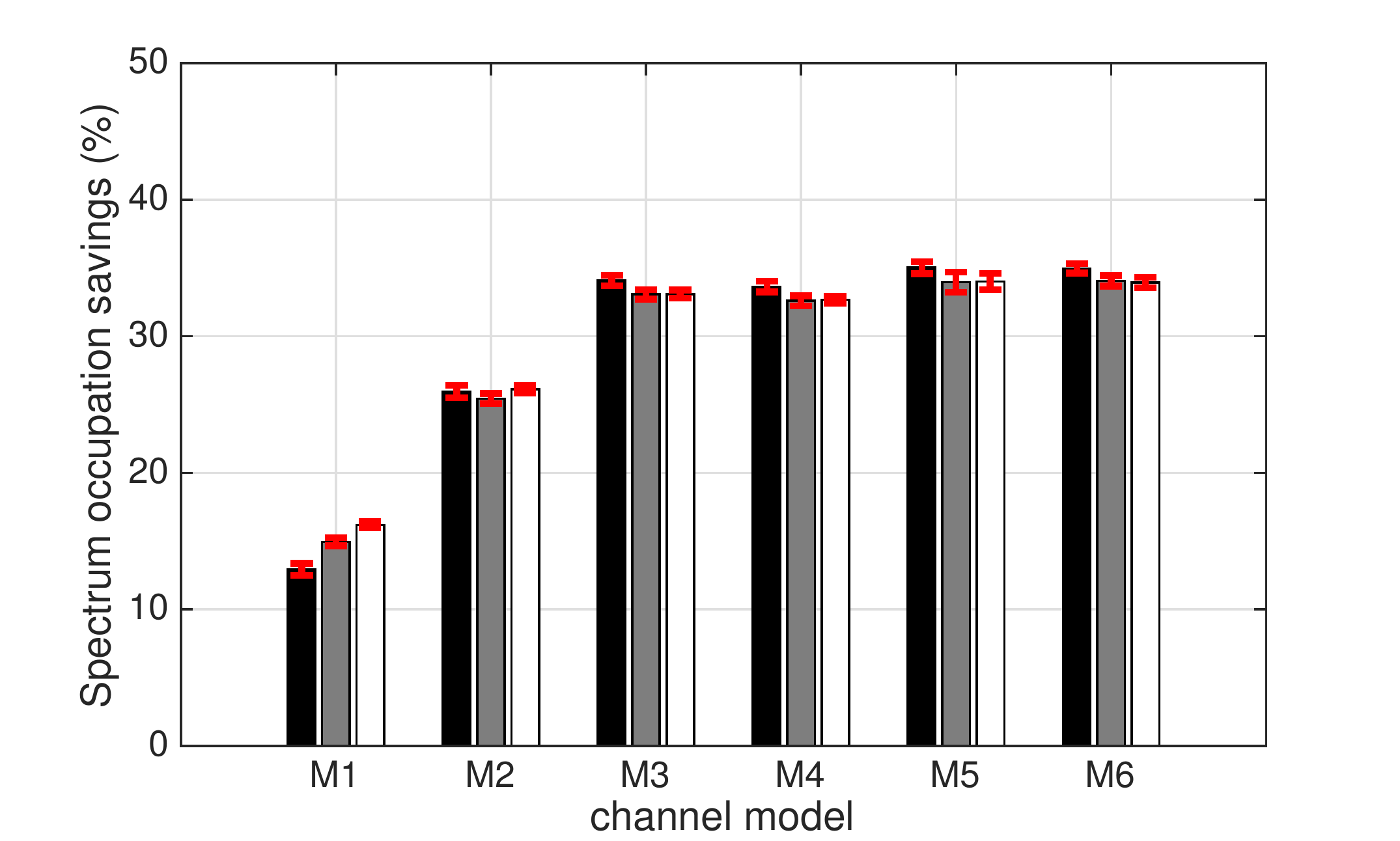}
\par\end{centering}
}\vspace*{-2mm}
\caption{Spectrum occupation savings: relative percentage of non-used spectrum
for with the CDMS-aided scheme, with respect to the amount of spectrum
used by the benchmark scheme; speed range {[}6,16{]} (black bars),
{[}9,24{]} (grey bars), {[}12,32{]} (white bars).}\label{R1_SPECTRUM_SAVINGS}
\vspace*{-2mm}
\end{figure*}

The same problem with M1 and M2 arises when considering the relative
gain in spectrum occupation, displayed in Figure~\ref{R1_SPECTRUM_SAVINGS}. In this case,
the mismatch is considerable.

The use of the deterministic models M1 and M2 (with the considered
link margin settings) tend to underestimate the possibility to spatially
reuse the spectrum, as the percentage spectrum occupation reduction
ranges (depending on the scenario, A or B, and the channel model,
M1 or M2) from 5\% to 21\%, while, when using channel models M3-M6,
it is between 31\% and 35\% for the different speed ranges.

The gain estimated with the simplistic models,
in this case, is completely unreliable. For instance, it can be up to six times less (with M1) than the what is estimated by the more accurate channel models.
This happens due to the effect explained in Figure~\ref{fig:PL}, i.e., the underestimation of the channel gain
difference between short ranges and long ranges, for both models M1
and M2. This underestimation turns into an overestimation of the interference
among potentially concurrent D2D links. As the RR scheduler takes
the estimated interference as an input, it tends to separate potentially
concurrent links more than actually necessary.\vspace{-2mm}

\section{Conclusion}\label{sec:conclusion}\vspace{-2mm}

In this work, we have studied the performance of a D2D-based traffic
offloading protocol for highly dynamic scenarios like vehicular environments.
We have evaluated its performance in terms of energy efficiency and
system spectral efficiency, by considering a selection of different, increasingly
detailed, wireless channel models. We have showed that the considered protocol
can achieve substantial improvements in terms of power consumption
and system-wise spectral efficiency (compared to a scheme that only
uses I2D communications). Specifically, for the vehicles traffic parameter
settings used in our study, we have observed a 35\% energy consumption
reduction gain and a 35\% reduction of spectrum use, with respect
to a benchmark scheme which uses classical I2D communications only.
In terms of spectral \emph{efficiency} (measured in bps/Hz), a 35\% reduction
of spectrum use corresponds to a 50\% increase.


Taking the channel model M6 (the most realistic channel model among
those considered in this work) as a reference, we have shown that
the simpler channel models, because they do not capture many aspects
of the radio propagation, are not able to consistently and reliably
evaluate the performance metrics. More specifically, we have seen
that the use of any of the models M1-M5, is either considerably unreliable
or, in the cases it provides an estimation of one of the considered
performance metrics (energy consumption and spectral occupancy) close
to the one provided by model M6, it fails to do so with the other
one. We have provided insights on the reasons of this behavior, pointing
out the major weaknesses of the considered channel models: the mismatch
of simple propagation loss formulas (in M1, M2) with respect to formulas
obtained through large measurements campaigns; the absence, in M1-M3,
of random fluctuations around the mean path loss due to shadowing
and small scale fading; and the lack of representation of the frequency
selectivity (in M1-M5). The overlooking of either of these aspects
prevents to correctly take into account important system parameters,
like the link margin and the FEC coding gain, which are quite important
for a system-level design and performance evaluation. Our analysis
and discussion, suggest that the performance evaluation of offloading
schemes based on a channel model, like M6, that captures the effects
of real-life propagation losses and channel frequency selectivity.
The use of accurate channel models, paired with a suitable physical
layer model, provides much more insights on the effects of physical
layer parameters with respect to the popular approach of using simplistic
models, and should play a significant role in the design of offloading
schemes, further optimized taking into account the above described
effects.\vspace*{-4mm}

\section*{Acknowledgement\vspace*{-2mm}
}

This work was partially funded by the EC under the H2020 REPLICATE
(691735), SoBigData (654024) and AUTOWARE (723909) projects.\vspace*{-3mm}

\bibliographystyle{elsarticle-num}

\vspace*{-2mm}

\section*{Appendix: Radio resource allocation for coexisting I2D and D2D links}

The ``full resource sharing'' approach \cite{Yang2017} builds on the idea of \emph{radio
resource set partitioning}. Conceptually, the procedure requires two
stages: first, the overall set of links to be scheduled (both I2D
and D2D links) is partitioned into subsets, called resource reuse
sets; second, each resource reuse set of links is assigned a subset
of the radio resources (the PRBs) available in a CI. In \cite{Yang2017},
both stages are performed in one-shot, having in input the transmit
power of each link\footnote{In \cite{Yang2017}, the transmit power is assumed to have a common
value for all I2D links, and another common value for all D2D links.}. The second stage also includes the possibility to prune some resource
reuse sets from the scheduling of the set of resources in order to
guarantee I2D communications (which have an higher priority).

Our RR scheduler is inspired by the scheduler proposed in \cite{Yang2017}
but differs from it in several aspects. Specifically, our implementation
copes with the more realistic assumptions of multiple cells, independent
power setting for each transmission, and frequency selective channels.
Furthermore, it guarantees that if a link is admitted to be scheduled
in a given CI, it will receive an amount of resources sufficient for
transmitting the whole desired content.

\begin{figure}[t]
\begin{centering}
\includegraphics[width=0.90\columnwidth]{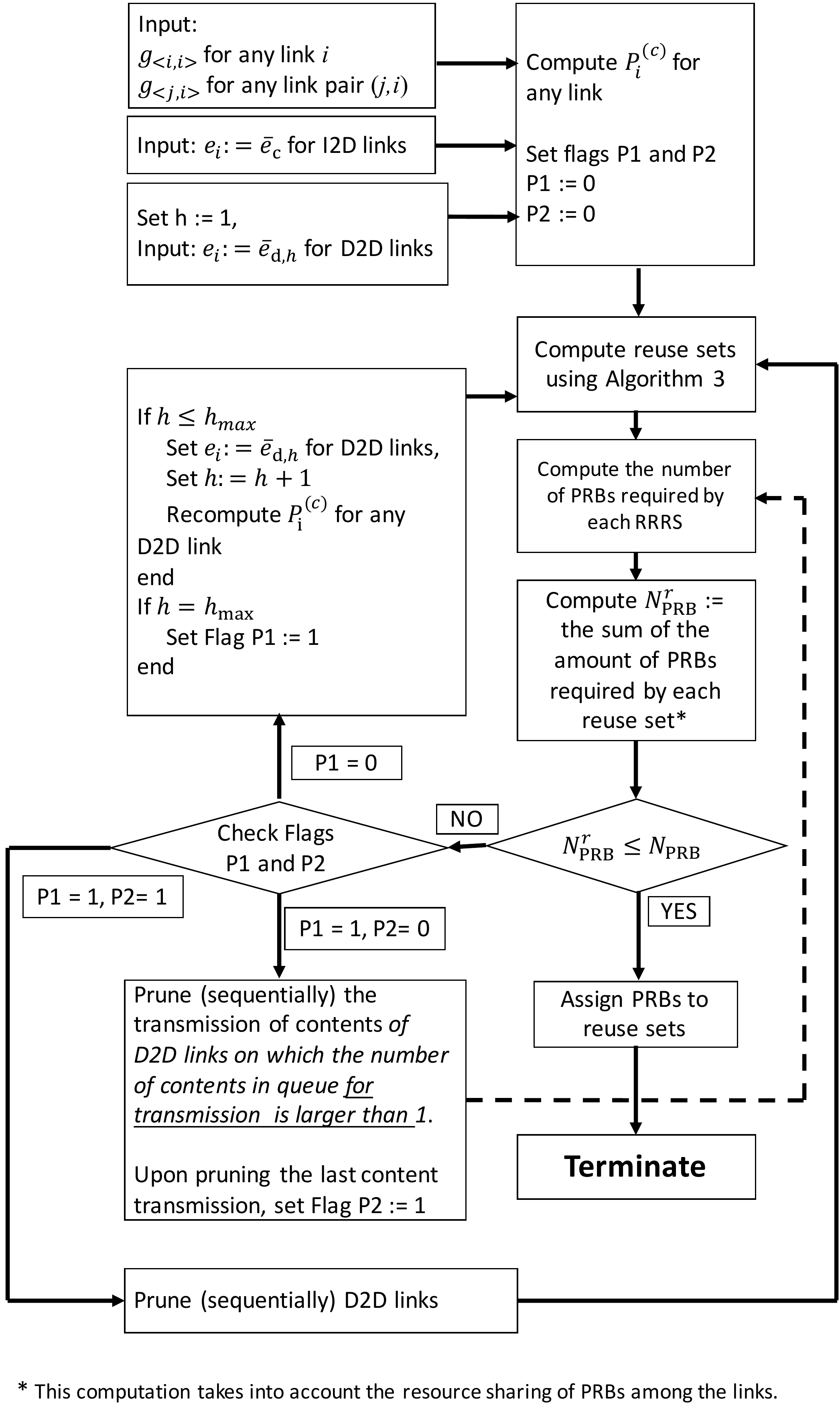}\vspace*{-4mm}
\par\end{centering}
\caption{Diagram of the considered radio resource allocation procedure.}
\label{fig:diagram}
\vspace*{-10mm}
\end{figure}

The whole scheduling procedure is sketched out in the diagram in Figure~\ref{fig:diagram}.
Here, for space reasons, we omit the details of the procedure, and
just highlight the most important features.

Rather then working at the link level only, we deal with the individual
I2D and D2D packet transmissions that the CDMS has established as
candidates for being scheduled. A single link may be required to transmit
one, \emph{or more}, packets during a CI. The scheduler inputs are:
all the nominal channel gains between any link pair $\left\langle j,i\right\rangle $;
for I2D links, a prescribed value of the target normalized nominal
information rate $e_{i}=\bar{e}_{\text{c}}$, equal for all the I2D
links; for D2D links, a set of candidate values for the target normalized
nominal information rate $\bar{\mathcal{E}}_{\text{d}}=\{\bar{e}_{\text{d},1},\ldots,\bar{e}_{\text{d},h_{\max}}\}$.
The computation of the per-subcarrier power levels, performed by the scheduler,
follows from Eq.~\eqref{eq:Power_per_subcarrier_dBm}. 

Now, if the modulation scheme in use has a maximum transmit spectral
efficiency, say $e$ bps/Hz, it makes no sense that the target normalized
information rate is set with a value larger than $e$, since the physical
transmission system will be in any case not able to transmit bits
with such a rate. Therefore, we set $\bar{e}_{\text{c}}$ to be the
maximum allowed spectral efficiency among the modulation formats in
use, and $\bar{\mathcal{E}}_{\text{d}}$ to be the set of values for
the transmit spectral efficiencies entailed, for example, by a set
of possible constellations that the D2D links can use. The rationale
behind this choice is as follows: first, for I2D links, which by definition
cannot share resources among them, the ideal thing to do is to use
as few PRBs as possible, in such a way to leave more room for the other
I2D communications. This is achieved by using the maximum transmit
spectral efficiency. However, to actually manage to communicate the
desired amount of information, a suitable power is required on each
subcarrier, and this power is determined by Eq.~\eqref{eq:Power_per_subcarrier_dBm}.
Second, for D2D links, which can share resources with other D2D links (and with
I2D links), since the resource reuse subsets will tend to favor the
coexistence of links with low cross-interference, the ideal thing
to do is to maximize the number of used resources, in such a way to
minimize the overall transmit power used to transmit a content\footnote{In general, in a radio communication, the larger the available bandwidth,
the lower the power required to obtain a desired capacity.}. For this reasons, the value of $e_{i}$ for D2D links, in the resource
allocation procedure, is initialized with the lowest value in $\bar{\mathcal{E}}_{\text{d}}$,
and the corresponding required power to support that rate is computed
according to Eq.~\eqref{eq:Power_per_subcarrier_dBm}. During the
procedure, after the computation of the radio resource reuse sets
(see below) and of the number of total required PRBs, $N_{\text{PRB}}^{r}$,
it may happen that the currently selected value of $e_{i}$ for D2D
links prevents to accomodate all the required transmissions, i.e.,
it results in $N_{\text{PRB}}^{r}>N_{\text{PRB}}$. In this case,
the scheduler iteratively increases the desired target nominal normalized
information rate $e_{i}$ for D2D links, it recomputes the resource
reuse sets and the corresponding number of total required PRBs, and
it re-checks if $N_{\text{PRB}}^{r}>N_{\text{PRB}}$. 
If no value of $e_{i}$ for D2D links, compatible with the requirement
$N_{\text{PRB}}^{r}\leq N_{\text{PRB}}$, is found, the procedure
starts to exclude D2D packets from those admitted for transmission
in the current CI, and then entire D2D links, until a subset of admissible
D2D links and D2D packets (each link can transmit one or more packets) to be transmitted, compatible with the requirement
$N_{\text{PRB}}^{r}\leq N_{\text{PRB}}$, is found.

The details of the computation of the resource reuse sets are summarized by Algorithm~\ref{algoRRRS},
which is based on \cite[Algoritm 1]{Yang2017}.
The following notation is required to interpret
Algorithm~\ref{algoRRRS}: let $S^{c}$ and $S^{d}$ be
the sets of I2D and D2D links among which the radio resources need
to be partitioned, with $S=S^{c}\cup S^{d}$. Let $S_{r}^{c}$ and
$S_{r}^{d}$ be the set of cellular and D2D links in the $r$-th radio
resource reuse set $S_{r}=S_{r}^{c}\cup S_{r}^{d}$ in the partition
$S=\uplus_{r}S_{r}$, which is the output of Algorithm~\ref{algoRRRS}.
We indicate with $b(i)$ the base station to which the receiver of
the I2D link $i$ is associated and with $\left\{ S_{1}^{c},\ldots,S_{N_{\text{BS}}}^{c}\right\} $
a partition of the link set $S^{c}$, where $i\in S_{b'}^{c}\text{ iff }b(i)=b'$,
or $S_{b'}^{c}=\left\{ i\in S^{c}\,|\,b(i)=b'\right\} .$ Finally,
we indicate with $\hat{P}_{r}(b,i)=\underset{j\in S_{b}^{c}}{\text{argmax}}\left(P_{j}^{(c)}g_{\left\langle j,i\right\rangle }\right)$
the maximum (nominal) interference that a D2D link $i$ can receive,
on a single subcarrier, from a I2D link which is in the same resource
reuse set as link $i$, and is handled by BS $b$. Finally, let $CL$
represent a running set of candidate links, initialized with the whole
set $S$ of links to be scheduled.\vspace*{2mm}

Algorithm~\ref{algoRRRS} results from \cite[Algoritm 1]{Yang2017}
replacing the minimum SNIR constraints \cite[Equations (2a) and (2b)]{Yang2017}
with\vspace*{2mm}
\begin{align}
{\scriptstyle \log_{2}\left(1+\frac{P_{i}^{(c)}g_{\left\langle i,i\right\rangle }}{\sigma_{c}^{2}+\sum_{j\in S_{r}^{d}}P_{j}^{(c)}g_{\left\langle j,i\right\rangle }}\right)\geq e_{i},\,\,\,\,\,\,\,\,\,\,\,\,\,\,\,\,\,\,\,\,\,\,\,\,\,\,\,\forall i\in S_{r}^{c}},\label{eq:MRC}\\
{\scriptstyle \log_{2}\left(1+\frac{P_{i}^{(c)}g_{\left\langle i,i\right\rangle }}{\sigma_{c}^{2}+\sum_{b\in\mathcal{B}}\hat{P}_{r}(b,i)+\sum_{j\in S_{r}^{d}\backslash\{i\}}P_{j}^{(c)}g_{\left\langle j,i\right\rangle }}\right)\geq e_{i}},\nonumber \\
\hphantom{{\scriptstyle \log_{2}\left(1+\frac{P_{i}^{(c)}g_{\left\langle i,i\right\rangle }}{\sigma_{c}^{2}+\sum_{j\in S_{r}^{d}}P_{j}^{(c)}g_{\left\langle j,i\right\rangle }}\right)\geq e_{i}},}{\scriptstyle \forall i\in S_{r}^{d}},\nonumber 
\end{align}
and the Resource Sharing Conditions (RSC) in \cite[Equations (9) and (17)]{Yang2017},
with RSC-1:\vspace*{2mm}
\begin{align}
{\scriptstyle {\displaystyle \sum_{j\in S^{d}\backslash\{i\}}}P_{j}^{(c)}g_{\left\langle j,i\right\rangle }<\frac{P_{i}^{(c)}g_{\left\langle i,i\right\rangle }}{(1+\left|S^{d}\right|\xi_{h})^{\frac{1}{\left|S^{d}\right|}}-1}-\sigma_{c}^{2}},\,\forall i\in S^{d} & ,\label{eq:RSC1}\\
{\scriptstyle \text{where }}{\scriptstyle \xi_{h}=\max_{i\in S^{d}}\frac{P_{i}^{(c)}g_{\left\langle i,i\right\rangle }}{w\mathcal{N}_{0}}},\nonumber 
\end{align}
and RSC-2:
\begin{align}
{\scriptscriptstyle \begin{cases}
{\scriptscriptstyle {\displaystyle \sum_{j\in S^{d}}}P_{j}^{(c)}g_{\left\langle j,i\right\rangle }\leq\frac{P_{i}^{(c)}g_{\left\langle i,i\right\rangle }}{(1+\left|S^{d}\right|\xi_{h})^{\frac{\left|S_{b(i)}^{c}\right|}{\left|S^{c}\right|+\left|S^{d}\right|}}-1}-\sigma_{c}^{2}}{\scriptstyle ,} & {\scriptstyle \hspace{-4mm}\forall i\in S^{c}}\\
{\scriptscriptstyle {\displaystyle \sum_{j\in S\backslash\{i\}}\hspace{-2mm}}P_{j}^{(c)}g_{\left\langle j,i\right\rangle }\leq\frac{P_{i}^{(c)}g_{\left\langle i,i\right\rangle }}{\left(1+\left|S^{d}\right|\xi_{h}\right){}^{\frac{1}{\left|S^{c}\right|+\left|S^{d}\right|}}-1}-\sigma_{c}^{2}}, & {\scriptstyle \hspace{-4mm}\forall i\in S^{d}}
\end{cases}},\label{eq:RSC2}\\
{\scriptstyle \text{where }}{\scriptscriptstyle \xi_{h}=\max\left(\max_{i\in S^{d}}\frac{P_{i}^{(w)}g_{\left\langle i,i\right\rangle }}{\sigma_{c}^{2}},\max_{i\in S^{c}}\frac{P_{i}^{(w)}g_{\left\langle i,i\right\rangle }}{\left|S^{d}\right|\sigma_{c}^{2}}\right)}\nonumber 
\end{align}
respectively.

Finally, in Algorithm~\ref{algoRRRS}, the term $\check{e}_{i}$
represents a lower bound on the nominal normalized information rate
achievable of link $i$, and is defined as
\begin{algorithm}[t]
\caption{{\small{}Radio Resource Set Partitioning}}
\label{algoRRRS}

\scriptsize{
\begin{enumerate}
\item $r=1$
\item \textbf{while} $CL\neq\emptyset$ \textbf{do}
\item \quad{}$S_{r}=CL$
\item \quad{}// Trim $S_{k}$ according to the set of inequalities \eqref{eq:MRC}
\item \quad{}compute $\check{e}_{i}$ $\forall i\in S_{r}$;
\item \quad{}find link $i^{*}\in S_{r}$ with minimum $\check{e}_{i}$;
\item \quad{}\textbf{while} $\check{e}_{i^{*}}<e_{i}$ \textbf{do}
\item \quad{}\quad{}remove $i$ from $S_{r}$;
\item \quad{}\quad{}compute $\check{e}_{i}$ $\forall i\in S_{r}$
\item \quad{}\quad{}find link $i^{*}\in S_{r}$ with minimum $\check{e}_{i}$;
\item \quad{}\textbf{end while}
\item \quad{}// Trim $S_{r}$ according to RSC1 and RSC2,
\item \quad{}find $S^{*}\subset S_{r}$ of the links not satisfying the
sets of inequalities \eqref{eq:RSC1} and \eqref{eq:RSC2};
\item \quad{}\textbf{while} $S^{*}\neq\emptyset$ \textbf{do}
\item \quad{}\quad{}compute $\check{e}_{i}$ $\forall i\in S^{*}$;
\item \quad{}\quad{}find link $i^{*}\in S^{*}$ with minimum $\check{e}_{i}$;
\item \quad{}\quad{}remove $i^{*}$ from $S_{r}$;
\item \quad{}\quad{}find $S^{*}\subset S_{r}$ of links not satisfying
the inequalities \eqref{eq:RSC1} and \eqref{eq:RSC2};
\item \quad{}\textbf{end while}
\item \quad{}remove link members of $S_{r}$ from $CL$ and set $k=k+1$;
\item \textbf{end while}
\end{enumerate}
}
\end{algorithm}
\[
{\scriptstyle \check{e}_{i}=\begin{cases}
\log_{2}\hspace{-1mm}\left(1\hspace{-1mm}+\hspace{-1mm}\frac{P_{i}^{(c)}g_{\left\langle i,i\right\rangle }}{\sigma_{c}^{2}+\sum_{j\in S_{r}^{d}}P_{j}^{(c)}g_{\left\langle j,i\right\rangle }}\right),\qquad\forall i\in S_{r}^{c},\\
\\
\log_{2}\hspace{-1mm}\left(1\hspace{-1mm}+\hspace{-1mm}\frac{P_{i}^{(c)}g_{\left\langle i,i\right\rangle }}{\sigma_{c}^{2}+\sum_{b\in\mathcal{B}}\hat{P}_{r}(b,i)+\sum_{j\in S_{r}^{d}\backslash\{i\}}P_{j}^{(c)}g_{\left\langle j,i\right\rangle }}\right),\\
\hphantom{\log_{2}\hspace{-1mm}\left(1\hspace{-1mm}+\hspace{-1mm}\frac{P_{i}^{(c)}g_{\left\langle i,i\right\rangle }}{\sigma_{c}^{2}+\sum_{j\in S_{r}^{d}}P_{j}^{(c)}g_{\left\langle j,i\right\rangle }}\right)}\hspace{10mm}\forall i\in S_{r}^{d}.
\end{cases}}
\]

\end{document}